\documentclass[a4paper,fleqn,usenatbib]{mnras}

\usepackage[T1]{fontenc}
\usepackage{ae,aecompl}

\usepackage{graphicx}	
\usepackage{amsmath}	
\usepackage{amssymb}	
\usepackage{hyperref}

\newcommand{\pin}{{\sc pinocchio}}

\title[Improving clustering with LPT]{Improving fast generation of halo catalogs with higher-order Lagrangian perturbation theory}

\author[Munari et al.]
 {Emiliano Munari$^{1,2,3}$\thanks{E-mails: munari@inaf.oats.it},
Pierluigi Monaco$^{1,2}$,
Emiliano Sefusatti$^{4}$,
Emanuele Castorina$^{5,6}$,\and
Faizan G. Mohammad$^{4}$,
Stefano Anselmi$^{7}$,
Stefano Borgani$^{1,2,8}$\\
$^1$ Dipartimento di Fisica - Sezione di Astronomia, Universit\`a di Trieste, via Tiepolo 11, I-34131 Trieste -- Italy\\ 
$^2$ INAF, Osservatorio Astronomico di Trieste, Via Tiepolo 11, I-34131 Trieste -- Italy \\
$^3$ Dark Cosmology Centre, Niels Bohr Institute, University of Copenhagen, Juliane Maries Vej 30, DK-2100 Copenhagen, Denmark\\ 
$^4$ INAF, Osservatorio Astronomico di Brera, Via Bianchi 46, I-23807 Merate (LC) -- Italy\\
$^5$ SISSA, International School for Advanced Studies, Via Bonomea 265, Trieste 34136, Italy\\
$^6$ Berkeley Center For Cosmological Physics,  Department of Physics and Lawrence Berkeley National Laboratory,\\ University of California, Berkeley, CA 94720, USA\\
$^7$ Department of Physics/CERCA/Institute for the Science of Origins, Case Western Reserve University, Cleveland, OH 44106-7079, USA\\
$^8$ INFN, Istituto Nazionale di Fisica Nucleare, Trieste -- Italy \\ 
}

\date{Accepted XXX. Received YYY; in original form ZZZ}

\pubyear{2015}

\begin{document}
\label{firstpage}
\pagerange{\pageref{firstpage}--\pageref{lastpage}}
\maketitle

\begin{abstract}
We present the latest version of {\pin}, a code that generates
catalogues of DM haloes in an approximate but fast way with respect to
an N--body simulation. This code version extends the computation of
particle and halo displacements up to 3rd-order Lagrangian
Perturbation Theory (LPT), in contrast with previous versions that
used Zeldovich approximation (ZA).

We run {\pin} on the same initial configuration of a reference N--body
simulation, so that the comparison extends to the object--by--object
level. We consider haloes at redshifts $0$ and $1$, using different
LPT orders either for halo construction - where displacements are
needed to decide particle accretion onto a halo or halo merging - or
to compute halo final positions.

We compare the clustering properties of {\pin} haloes with those from
the simulation by computing the power spectrum and 2-point correlation
function (2PCF) in real and redshift space (monopole and quadrupole),
the bispectrum and the phase difference of halo distributions. We find
that 2LPT and 3LPT give noticeable improvement. 3LPT provides the
  best agreement with N--body when it is used to displace haloes,
  while 2LPT gives better results for constructing haloes. At
the highest orders, linear bias is typically recovered at a few per
cent level.

In Fourier space and using 3LPT for halo displacements, the halo power
spectrum is recovered to within 10 per cent up to 
$k_{max}\sim0.5\ h/$Mpc. The results presented in this
  paper have interesting implications for the generation of large
  ensemble of mock surveys aimed at accurately compute covariance
  matrices for clustering statistics.

\end{abstract}


\section{Introduction}

Many galaxy surveys that have started or are planned in the forthcoming
years will map increasingly larger regions of the Universe, such
as 
\clearpage\noindent
DES\footnote{http://www.darkenergysurvey.org/} \citep[Dark Energy Survey][]{frieman2013}, 
DESI\footnote{http://desi.lbl.gov/} \citep[Dark Energy Spectroscopic Instrument][]{schlegel2011,levi2013}, 
eBOSS\footnote{http://www.sdss.org/surveys/eboss/} (Extended Baryon Oscillation Spectroscopic Survey), 
LSST\footnote{http://www.lsst.org/} \citep[Large Synoptic Survey Telescope][]{abell2009}, 
Euclid\footnote{http://sci.esa.int/euclid/} \citep{laureijs2011}, 
WFIRST\footnote{http://wfirst.gsfc.nasa.gov/} \citep[Wide-Field Infrared Survey Telescope][]{green2012}
 and SKA\footnote{http://www.skatelescope.org/} surveys.

With the number of observed galaxies growing to billions, these
observations will constrain cosmological parameters with an accuracy
that will completely depend on the control that we have on the
systematics and covariances of observables. These are dominated by the
interplay between large-scale fluctuations and the non-Gaussian
properties of the galaxy distribution \citep[see,
  e.g.][]{hamilton2006,deputter2012}, by the bias with which galaxies
trace the underlying mass field, and by our knowledge of the sample
volume and selection function. The most effective way to take full
control of these quantities is to simulate a large number of galaxy
catalogues (mock catalogues) that reproduce in a realistic way the
actual survey. When dealing with the clustering of galaxies, a very
large (several thousands or more) number of realizations of the survey
are necessary to robustly estimate the covariance of the clustering
measurements \citep[see,
  e.g.][]{hartlap2007,mohammed2014,niklasgrieb2015,oconnel2015,percival2014,sunayama2015,paz2015}.

In fact, few realizations make the estimate of the covariance matrix
noisy. The estimate of cosmological parameters further involves the
inversion of the covariance matrix, that is very sensitive to the
level of noise. For such a large number of realizations, a program
simply based on N--body simulations would be
unfeasible. \cite{pope2008} introduced the so-called \emph{shrinkage
  technique} that allows to combine unbiased, high variance estimates
and biased, low variance estimates of the covariance matrix, allowing
to achieve very accurate estimates of it. Few N-body simulations are
therefore enough to provide the unbiased, high variance estimate of
the covariance matrix. Conversely, in order to generate synthetic halo
catalogs, it is possible to exploit analytic approximations to the
non-linear growth of structure to obtain a good approximation of the
density field on large scales and of the Dark Matter (DM) halo
distribution, as pioneered by, e.g., \cite{coles1993} and
\cite{borgani1995}, to obtain the biased, low variance estimates.

Many different methods have been proposed in the past years to
generate realizations of DM haloes in an approximate but fast way,
like e.g. {\pin} \citep[][hereafter M02 and M13, respectively
]{monaco2002,monaco2013}, PTHalos
\citep{scoccimarro2002,manera2013,manera2015} COLA
\citep{tassev2013,izard2015,koda2015,howlett2015,tassev2015}, PATCHY
\citep{kitaura2014,kitaura2015}, HALOgen \citep{avila2014}, EZmock
\citep{chuang2015}, QPM \citep{white2014}, and FastPM
\citep{feng2016}. A comparison among these methods is presented in the
comparison paper by \cite{nIFTy} (nIFTy, hereafter), where catalogues
realized with these techniques are compared with a reference N-body
simulation.

In this paper we present the latest version of the code {\pin}
(PINpointing Orbit Crossing Collapsed HIerarchical Objects), showing
how the accuracy with which clustering is reproduced increases when
using higher orders (up to the 3rd) of Lagrangian Perturbation Theory
(LPT hereafter) in the prediction of DM halo displacements.  {\pin},
first presented in M02 and \cite{taffoni2002}, with successive
modifications presented in M13, is a fast approximate tool for
generating DM halo catalogues, light cones and halo merger histories
starting from a set of initial conditions identical to those used for
N-body simulations. In M13 the code was redesigned to be fully
parallel and suitable for large cosmological volumes.  In this code,
described in some detail below and in two appendices, particle collapse
times are computed using ellipsoidal collapse, whose solution is
analytically obtained applying third-order LPT (3LPT) to a homogeneous
ellipsoid.  This allows to reconstruct, with good accuracy, the
Lagrangian patches that are going to collapse into a DM halo at a
later time. However, in previous versions of the code the
displacements from the Lagrangian space to the Eulerian configuration
at a given time were performed using the Zeldovich approximation (ZA),
the first-order term of LPT.  This reflects into a significant loss of
power already at $k\sim0.1\ h/{\rm Mpc}$ \citep{monaco2013}. The code
has thus been extended to use 2LPT or 3LPT for the
displacements. Results obtained with 2LPT version were already shown
in the nIFTy mock comparison paper. In what follows we will show and
quantify the improvement brought by this extension to the ability of
{\pin} to predict the clustering of DM haloes.

The paper is structured as follows. In Section~\ref{section:methods}
we present the N--body simulation and the {\pin} code used for this
paper, and the configurations of the catalogues that will be analysed.
In Section~\ref{sect: object by object} we present a comparison
between N--body and {\pin} haloes on an object--by--object basis.
Section~\ref{sect: clustering} presents the results of the clustering
analysis, and in Section~\ref{section: conclusions} we present the
conclusions. In Appendix~\ref{Appendix: Implementation and
  performance}, information on the implementation and on the
performance of the code are presented, while Appendix~\ref{app:calib}
gives more details on the calibration process.

The adopted cosmological parameters are the following: $\Omega_m =
0.25$, $\Omega_{\Lambda} = 0.75$, $\Omega_b = 0.044$, $h = 0.7$, and
$\sigma_8 = 0.8$.
{\pin}\footnote{http://adlibitum.oats.inaf.it/monaco/Homepage/Pinocchio/index.html}
is distributed under a GNU-GPL license.

\section{Methods}
\label{section:methods}

\subsection{N-body Simulation}
\label{section:nbody}

To assess the accuracy of {\pin} in reproducing the clustering of DM
haloes, we used an N-body simulation run with the {\sc gadget 3}
\citep{springel2005} code.  The box represents a volume of 1024 $Mpc/h$
of side, sampled with $1024^3$ particles. With this resolution, and
with the cosmological parameters reported above, the particle mass is
$6.94 \times 10^{10}\ {\rm M}_\odot/h$. A Plummer-equivalent
gravitational softening of $1/50$ of the mean inter-particle distance
was used.  Haloes were identified using a standard friends-of-friends
(FOF) algorithm, with a constant linking length equal to $0.2$ times
the inter-particle distance.  We consider as faithfully reconstructed
haloes those with $>100$ particles.

For the analysis on the accuracy of the reconstruction of haloes, we
make use of the same kind of simulation described above, run with
$512^3$ particles in a $512$ Mpc/h side box. The particle mass is the
same as in the bigger run.

\subsection{\pin}
\label{section:pinocchio}

{\pin} has been presented in M02, while an updated version (V3.0) was
presented in M13. We present here results from V4.0, a version that
has been completely re-written in the C-language and adapted to run on
massively parallel HPC infrastructures. A technical presentation and a
resolution test are presented in Appendix~\ref{Appendix:
  Implementation and performance}.

The algorithm behind the {\pin} code works as follows.  A linear
density field on a regular grid in Lagrangian space is generated in
the same way initial conditions are generated for an N--body
simulation. The density field is then smoothed on a set of scales, and
the collapse time for each particle at each scale is computed by
adopting the ellipsoidal collapse model based on LPT
\citep{monaco1995,monaco1997}.  Collapse here is defined as the event
of orbit crossing, or collapse of the ellipsoid on the first axis.
For each particle, the earliest collapse time (considering all
smoothing scales) is chosen as collapse time of the particle.  At this
time the particle is expected to become part of a DM halo or of the
filamentary network that connects the haloes.  In order to group
collapsed particles into DM haloes (selecting out the filamentary
network) and to construct halo merger histories, an algorithm (called
\emph{fragmentation}) that mimics the hierarchical process of
accretion of matter and merging of haloes, is applied. This algorithm
is already described in the papers cited above\footnote{
    small improvement is worth to be reported. Filament particles can
    be accreted when a neighbouring particle is accreted. Previous
    versions of the code neglected to check whether the filament
    particle that is accreted satisfies the accretion condition. We
    have added this condition, thus removing an anomalous growth of
    massive haloes, visible at the few percent level in the mass.};
a further improvement of V4.0 is an algorithm for on-the-fly
  reconstruction of the past light cone, that will be described and
  tested in a future paper.  The two main events of accretion of a
particle onto a halo and merging of two haloes are decided on the
basis of the following criterion: the two objects (particle-halo or
halo-halo) are displaced from the Lagrangian space to their expected
Eulerian position at the time considered, and accretion or
  merging take place if their distance $d$ is below a threshold
  $d_{\rm thr}$ that depends on the Lagrangian radius $R$ of the
  largest object.  This procedure implies (i) the use of LPT to
displace the two objects, (ii) free parameters to set the threshold.
Notably, the displacement of a halo is computed as the average
displacement of all particles that belong to it. This algorithm is
characterized by continuous time sampling, so the catalogue can be
output at any time and there is no need to output a large number of
``snapshots'', giving positions and velocities of all particles, to
reconstruct merger histories or the past light cone.  When a catalogue
is written, each halo is displaced from its Lagrangian position to its
Eulerian position at the desired redshift by applying LPT.

As a matter of fact, LPT displacements are performed in two different
occasions, during the construction of haloes and when computing the
position of haloes at catalogue output. It is convenient to leave freedom
to use different LPT orders for the two cases. This allows to test the
effect of increasing the LPT order for displacements while producing
exactly the same catalogue of haloes, and the improvement given by higher
LPT orders in halo construction when the displacements are computed at
the same order.  In the following we will always specify what LPT
order is used (ZA, 2LPT, 3LPT), separately for halo construction and
halo displacement, in this order.

The runs produced in this paper have the same cosmology, box size,
number of particles and random seeds (i.e. large-scale structures) as
the N-body simulations described in Section~\ref{section:nbody}.

We will make use of two different setups, already described
  in Section~\ref{section:nbody}. The smaller simulation, with $512^3$
  particles in a 512 Mpc/h side box at $z=0$, will be used for the
  halo--by--halo comparison, while the larger simulation, consisting
  of $1024^3$ particles in a 1024 Mpc/h side box, will be used at
  $z=0$ and 1 for the clustering analysis. Particle mass is $6.9 \cdot
  10^{10} M_\odot h^{-1}$ in both cases.

\subsubsection{Calibration of the mass function}
\label{section:mf}

We start by recalling that the mass function of dark matter haloes is
approximately ``universal'', meaning that mass functions at all
redshifts and for all cosmological parameters lie on the same
relation, when the adimensional quantity $(M^2/\bar\rho)n(M)$
($n(M)dM$ being the number density of haloes of mass between $M$ and
$M+dM$) is shown as a function of $\nu=\delta_c/\sigma(M)$, with
$\sigma(M)$ being the mass variance at the scale $M$ and
$\delta_c=1.686$ the usual density contrast for spherical top-hat
collapse. Recent results \citep[e.g.][]{tinker2008,crocce2010}
show that the mass function obtained from N-body simulations violates
universality, but these violations are relatively small and depend (as
the mass function itself) on how haloes are extracted from the
simulation, and on the choice of the overdensity \citep{despali2015}.

In the {\pin} code, the expression for the threshold
  distance $d_{thr}$ that determines accretion and merging includes
  free parameters. These must be calibrated by requiring that the halo
  mass function reproduces that of N-body simulations in a wide range
  of mass resolutions and redshifts. As long as these parameters are
  formulated in a cosmology-independent way, their calibration is
  performed once for all. We verified that the parameterizations of
  $d_{thr}$ used both in M02 and M13 were unable to produce a truly
  universal mass function, or a midly non-universal one in a way that
  resembles numerical results. We then reformulated the
  parametrization of $d_{\rm thr}$.

If particle displacements were as accurate as the N-body ones, we
would need only one parameter, setting the average density of the
reconstructed halo. Assuming for the moment that an overdensity of
$200$ (with respect to the mean density) corresponds to a virialized
halo, we could use, for both accretion and merging:

\begin{equation}
d_{\rm thr} = R / ^3\sqrt{200} = f_{200} R\, .
\label{eq:calib1}
\end{equation}

\noindent
 But LPT displacements are not accurate enough to use this
  simple formula. Also, the inaccuracy of displacements grows with the
  level of non-linearity, that is measured by $D(t)\sigma$, the
  standard deviation of unsmoothed density at the time $t$. So we
  expect, compared with simulations, a slower growth of haloes at
  later times. Conversely, what matters in Equation~\ref{eq:calib1} is
  the error in the displacement relative to the size $R$ of the halo,
  so that a given absolute value of the error is less relevant for the
  larger and more massive haloes. As a consequence, simply increasing
  the value of $f_{200}$ leads to an anomalous growth of massive
  haloes, more marked at later times, contrary to the expectation
  mentioned above. We then used this parametrization:

\begin{equation}
d_{\rm thr}^2 = (f R^{e})^2 + (f_{200} R)^2\, .
\label{eq:calib2}
\end{equation}

\noindent
 Here $f$ and $e$ are two free parameters; $f$ sets the
  threshold distance for small haloes, $e$ the velocity at which
  $d_{\rm thr}$ tends to $f_{200}R$ at large $R$. At the same time
  $f_{200}= 1/^3\sqrt{200}\simeq 0.171$ is held fixed. As in previous
  parametrizations, $f$ could have been chosen to be different for
  accretion ($f_a$) and merging ($f_m$). We tested this possibility,
  and noticed that $f_m$ and $e$ are degenerate, so the two-parameter
  formulation is adequate.

 This simple parametrization allows us to obtain a nearly
  universal mass function, to a better level than previous ones. But,
  as expected, the mass of haloes grows in time less fast than in
  simulations, resulting in an underestimate of the mass function at
  late times. This is illustrated to a higher level of detail in
  Appendix~\ref{app:calib}. To compensate for this we resort to the
  following parametrization:

\begin{equation}
d_{\rm thr}^2 = \left\{ 
\begin{array}{ll}
(f R^{e})^2 + (f_{200} R)^2 & D\sigma\le D\sigma_0\\
\{f R^{e} [1+s_{m,a}(D\sigma-D\sigma_0)]\}^2 + (f_{200} R)^2 & D\sigma>D\sigma_0\\
\end{array}
\right.
\label{eq:calib3}
\end{equation}

\noindent
 The value of $D\sigma_0$ is directly obtained from runs
  performed using the two-parameter setting of
  equation~\ref{eq:calib2}, as explained in Appendix~\ref{app:calib}.
  In this case it is necessary to use different values of the $s$
  parameter for accretion and merging, $s_m$ and $s_a$.

 We found best-fit parameters, for the three LPT orders, in
  two cases. We required a best fit of the mass function of our
  $1024^3$ simulation, and we required to best fit a universal
  analytic mass function, based on many sets of runs. This second
  procedure is described in the Appendix~\ref{app:calib}. The fitting
  procedure for the simulation is simple: the first two parameters $f$
  and $e$ are obtained by fitting the N-body mass function at a time
  near $D\sigma=D\sigma_0$; $f$ mostly determines the normalization of
  the mass function, $e$ its slope. Then $s_a$ is chosen to obtain a
  satisfactory normalization at lower redshifts, while $s_m$ is tuned
  to correct for differences in the slope.

\begin{figure}
  \centering
  \includegraphics[width=\columnwidth]{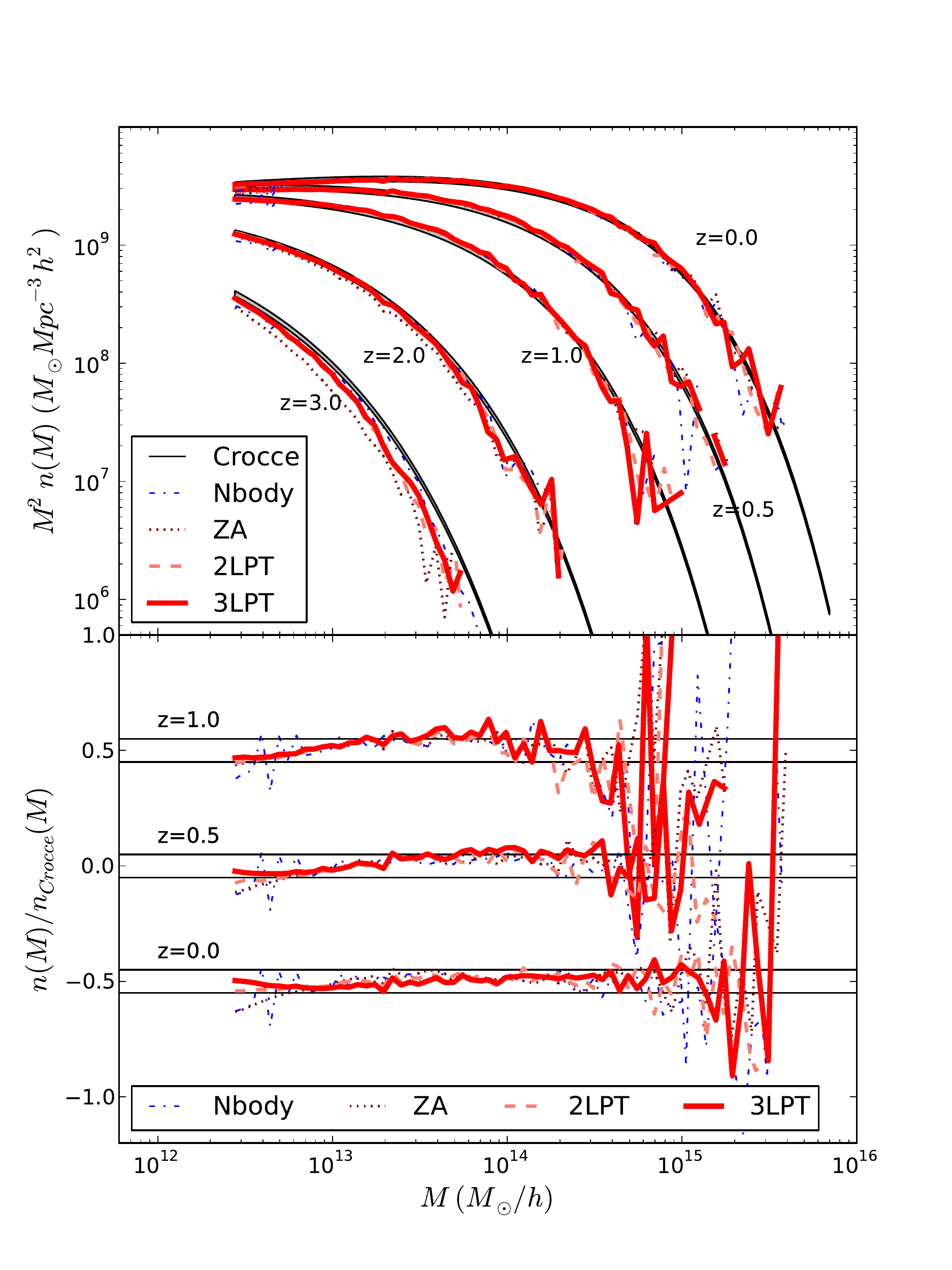}
  \caption{\label{fig:calibratedMF} Mass function of the
      N-body simulation (thin blue dot-dashed lines) and {\pin} using
      the three LPT orders to construct haloes, at various redshifts:
      ZA (dotted brown lines), 2LPT (dashed pink lines) and 3LPT
      (thick red solid lines). The upper panel shows the quantity
      $M^2n(M)$ as a function of $M$, black thick lines give the
      (non-universal) analytic fit of Crocce et al. (2010). The lower
      panel shows the residuals with respect to the analytic fit for
      $z=0$, $0.5$ and $1$, conveniently displaced. The horizontal
      black lines give the $\pm5$ per cent region around the Crocce
      fit.}
\end{figure}

 Figure~\ref{fig:calibratedMF} shows the resulting mass
  functions. Upper panels show the quantity $M^2 n(M)$ for five
  redshifts, while the lower panel shows residuals from the
  (non-universal) analytic fit of \cite{crocce2010}, that represents
  well the numerical mass function to within $\sim5$ per cent, for
  $z=1$, $0.5$ and $0$. Masses of FoF haloes, $M_{\rm FOF}$, have been
  corrected as in \citep{warren2006}: $M_{\rm FOF} = M_{\rm part}
  \cdot N \cdot (1 - N^{-0.6})$ (where the halo is made of $N$
  particles of mass $M_{\rm part}$), in order to avoid a known bias of
  FoF haloes sampled by few particles. From $z=0$ to $z=1$, the
  agreement is good at the $\sim1-2$ per cent level at small masses,
  for the three LPT orders, while at large masses sampling noise
  becomes larger and the agreement remains good within this noise. At
  large masses, correlation in the shot noise of mass functions is
  clearly visible, especially at $z=1$. Table~\ref{table:parameters}
  gives the best-fit parameters for the three LPT orders. At this
  resolution, a relatively high value of the $s_a$ parameter allows to
  reproduce the non-universality of the mass function found by
  \cite{crocce2010}. At $z\sim3$ the agreement worsens considerably,
  especially for ZA that is found to underestimate the numerical mass
  function also at $z=2$. We show in the Appendix~\ref{app:calib} that
  this is mostly an effect of resolution.

\begin{table}
\centering
\begin{tabular}{lrrr}
\hline
Parameter & ZA value & 2LPT value & 3LPT value\\
\hline
$f$         &  0.495 &  0.475 &  0.445\\
$e$         &  0.852 &  0.780 &  0.755\\
$s_a$       &  0.500 &  0.650 &  0.700\\
$s_m$       & -0.075 & -0.020 &  0.000\\
$D\sigma_0$ &  1.7   &  0.1.5 &  1.2  \\
\hline \hline 
\end{tabular}
\caption{\label{table:parameters} Adopted values of the parameters of
  eq. \ref{eq:calib3} for the calibration of the mass function against
  the N-body simulation.}
\end{table}

\subsection{Catalogue selection}
\label{sect: catalogues}

In this paper we will test the accuracy of {\pin} in reproducing the
results of N-body simulations run on the same initial conditions, both
on a halo--by--halo basis and by checking the accuracy of the
clustering statistics.

Catalogues for clustering analysis (based on the $1024^3$ setup)
  are defined as follows.  From the full catalogues obtained, at $z=0$
  and $z=1$, using both {\sc gadget} and {\pin}, we define two sets of
  catalogues by selecting the haloes with at least 100 or 500
  particles, corresponding to a selection of haloes more massive than
  $6.9 \cdot 10^{12} M_\odot h^{-1}$ and $3.4 \cdot 10^{13} M_\odot
  h^{-1}$, respectively. Before applying these cuts, the Warren
  correction \citep{warren2006} on the number of particles of FoF
  haloes is applied. These mass limits are set at the smallest FOF
halo mass where we are fully confident that the reconstruction is
correct and a larger mass where we still have sufficient statistics at
$z=1$. We want to stress that the lower mass cut is motivated by the
accuracy of N-body simulations, but {\pin} haloes, being built with a
semi-analytic approach, are considered reliable as long as the mass
function is well recovered.


\section{Object by object comparison}
\label{sect: object by object}

A thorough investigation of the ability of {\pin} to recover FOF haloes
from an N-body simulation run on the same seeds was presented in
M02.  In that paper halo matching was decided on the
basis of the number of particles in common between N-body and {\pin} haloes.
Our aim here is not to repeat the same detailed analysis (the main algorithm
has not changed) but to assess how the increased LPT order impacts on
group reconstruction.  We then adopt a simpler and faster criterion
for halo matching, 
and apply it to the smaller and more manageable $512^3$
simulation (see Section~\ref{section:nbody}) at $z=0$.
 
For each halo produced by the {\pin} algorithm we consider the first
particle that collapsed along the main progenitor branch, thus
providing the earliest seed for the construction of the halo. Since it
is likely to find this particle close to the bottom of the potential
well of the N-body halo, if the corresponding particle in the N-body
simulation belongs to a FOF halo then we match this {\pin} halo with
the FOF one.

\begin{figure}
  \centering
  \includegraphics[width=\columnwidth]{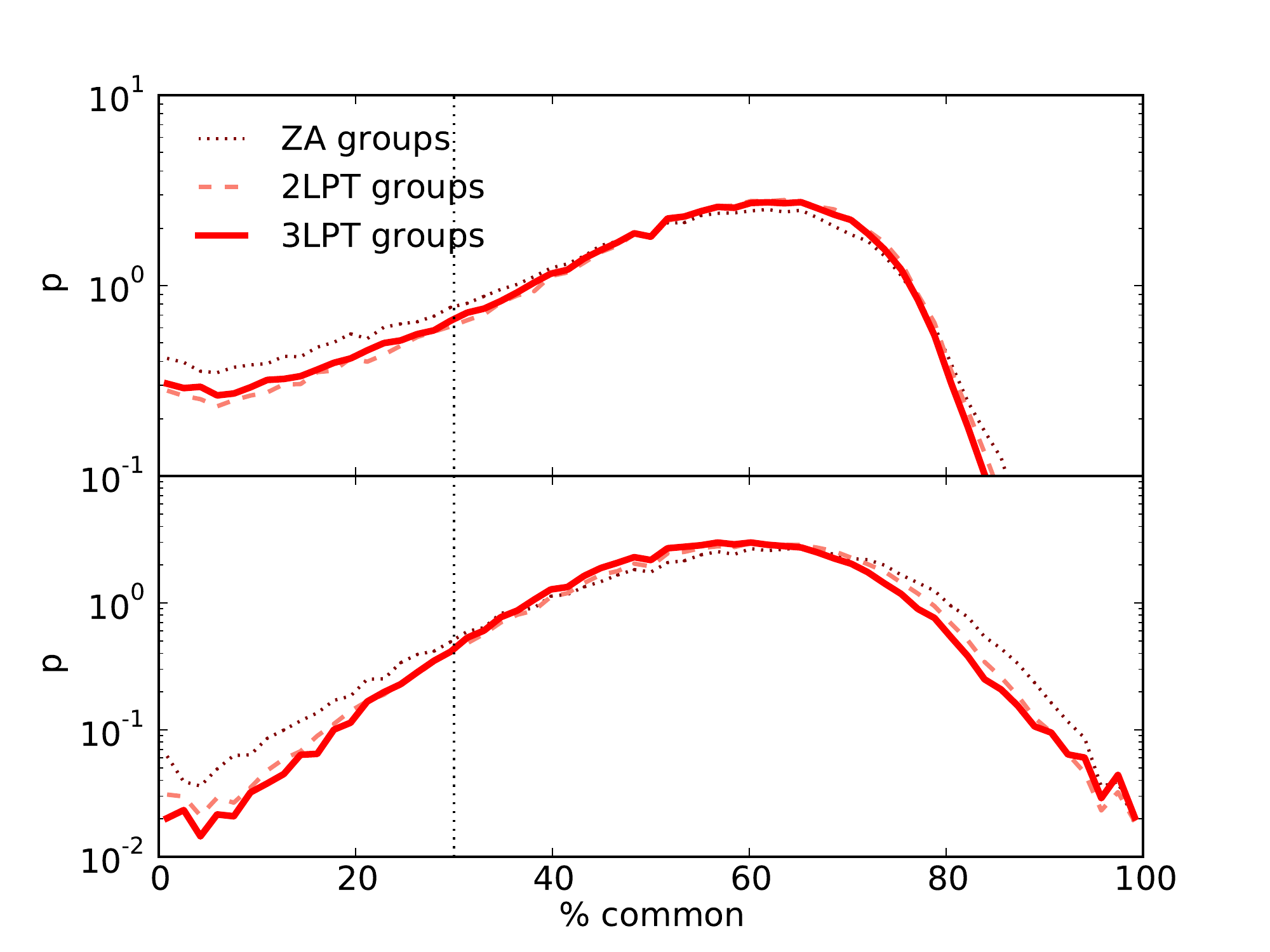}
  \caption{\label{fig: common} Distribution of the fraction of the
    particles that are in common between matched halo pairs, using
    different LPT orders in halo construction, as indicated in the label. The
    number of particles in common is normalized by the total number of
    particles of the {\pin} halo (\emph{top panel}) or by the total
    number of particles of the FOF halo (\emph{bottom panel}). In both
    panels, the distributions are normalized so that the integral of
    each one is unity. The vertical dotted line locates the threshold
    identifying the ``clean'' matched haloes.}
\end{figure}

Considering each matched halo pair, the number of particles in common
between the two haloes is recorded. Such number can be divided by the
total number of particles of the {\pin} halo or of the FOF halo.
Fig. \ref{fig: common} shows the distribution, for all matched halo
pairs, of the fraction of common particles with respect to the {\pin}
halo mass (upper panel) and with respect to the FOF halo mass (lower
panel).  Since the timing of the merging of two haloes into a bigger
one may not be identical in {\pin} and in simulation, it can happen
to match a halo before a merging with one after the corresponding
merging, or vice versa. This, together with possible mis-matches due
to numerics, justifies the tail to low values of these distributions.
Loosely following M02, we define ``cleanly matched''
haloes the pairs that have both fractions above $30$ per cent. This relatively
low value allows to obtain fraction of matched haloes in line with the
detailed results of the 2002 paper.  In Table \ref{tab: CMH} we report
the number of cleanly matched haloes found in the different runs, where
haloes are constructed with different LPT orders, as well as the total
number of haloes in {\pin} catalogues and in simulations and the
relative fractions.  Overall, $\sim70$ per cent of haloes are cleanly
reconstructed, the number being dominated by the smallest masses where
the reconstruction is less accurate.

\begin{table}
\centering
\begin{tabular}{llll}
\hline
\shortstack{Halo\\construction\\order} & N CMH & \shortstack{N in {\pin}\\ (\% of CMH)} & \shortstack{N in simulation\\ (\% of CMH)}  \\
\hline 
ZA & 134869 & 192819 (69.9\%) & 222118 (60.7\%)\\
2LPT & 146410 & 201019 (72.8\%) & 222118 (65.9\%)\\
3LPT & 144055 & 204687 (70.4\%) & 222118 (64.9\%)\\
\hline \hline 
\end{tabular}
\caption{\label{tab: CMH} Information on the statistics of cleanly
  matched haloes (CMH). In the first column the order with which
  haloes are constructed that identifies the catalogue used
  (independently of the displacement), in the second column the number
  of CMH, in the third column the total number of haloes in the full
  {\pin} catalogue and in parenthesis the percentage of the CMH, and
  in the fourth column the total number of haloes in the full FOF
  catalogue and in parenthesis the percentage of the CMH. Catalogues
  of haloes constructed with the same order of LPT but displaced with
  different orders of LPT provide the same number of matched haloes
  and are therefore not shown here.  }
\end{table}

\begin{figure}
  \centering
  \includegraphics[width=\columnwidth]{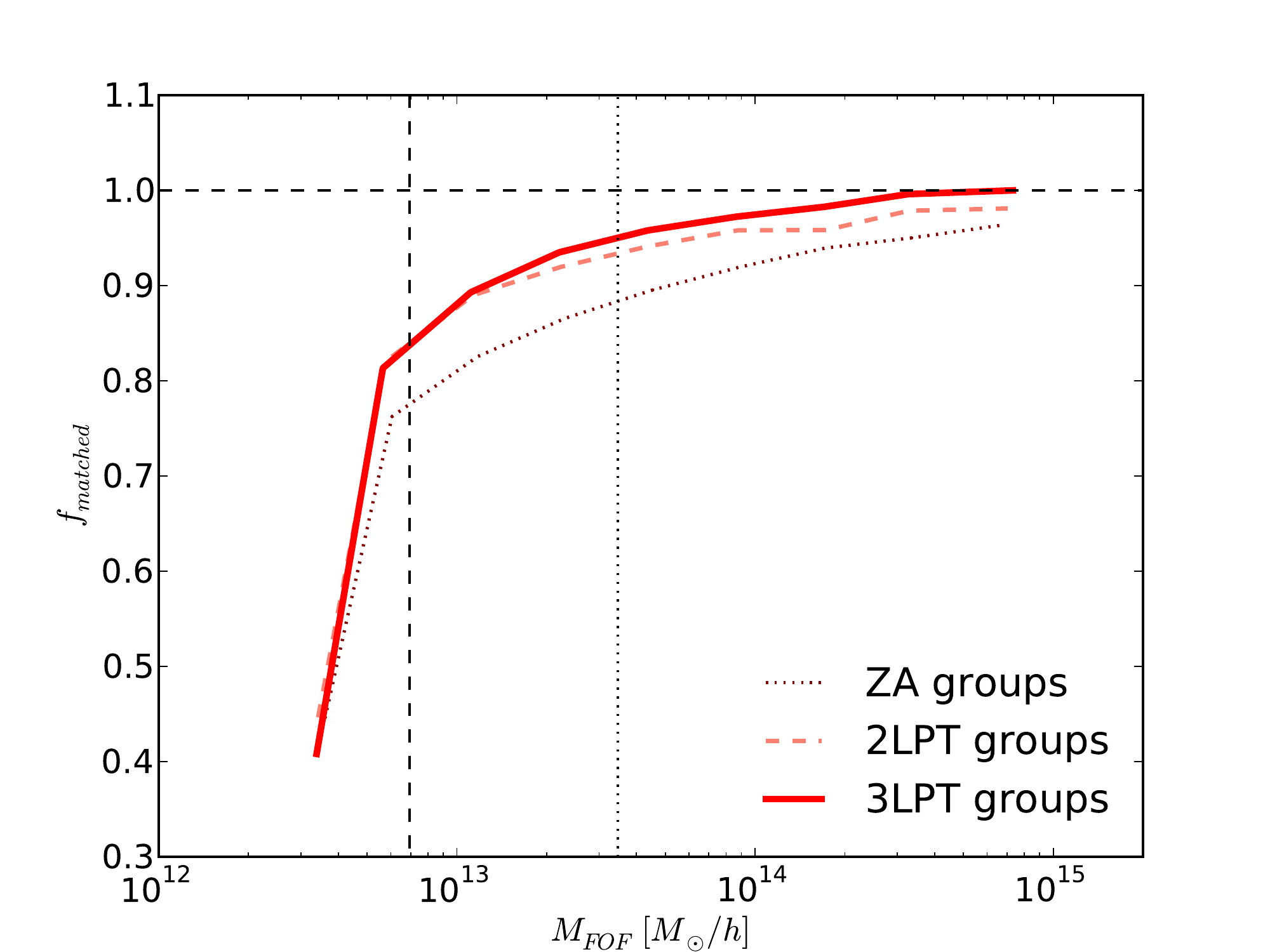}
  \caption{\label{fig: frac} Fraction of cleanly matched haloes with
    respect to the total number of haloes in {\pin} catalogues as a
    function of FOF halo mass. Different colours refer to different
    runs where haloes are built via ZA, 2LPT or 3LPT (see legend). The
    vertical dashed line identifies the mass corresponding to 100
    particles while the vertical dotted line locates the mass
    corresponding to 500 particles.}
\end{figure}

In Fig. \ref{fig: frac} the fraction of cleanly matched haloes (with
respect to the total number of haloes in {\pin} catalogues) is shown as
a function of FoF halo mass. In this plot we give results for haloes
starting from 32 particles, to show the degradation taking place below
the 100 particles limit. The higher order halo constructions (2LPT and
3LPT) perform better than ZA above $\sim5 \cdot 10^{12} M_\odot$, where
resolution is good enough to reconstruct the haloes.  The results for
ZA are very similar with what was found in M02.  It is
worth stressing again that the fall of this fraction is not
necessarily a sign of inaccuracy of {\pin}, but is surely due, at
least in part, to the difficulty of recognising simulated haloes
sampled by few particles.

\begin{figure}
  \centering
  \includegraphics[width=\columnwidth]{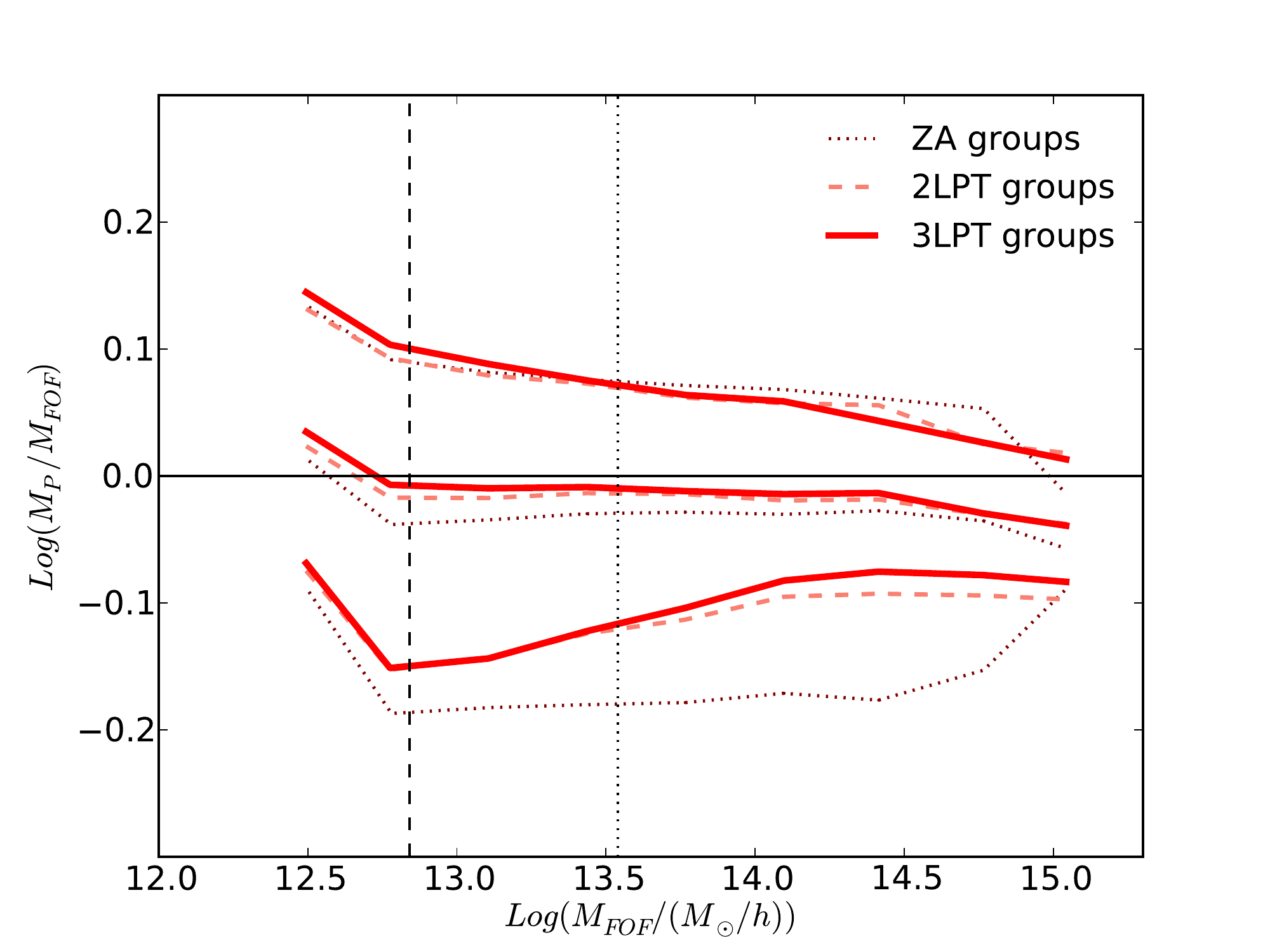}
  \caption{\label{fig: massscatter} Relative accuracy of mass
    reconstruction for cleanly matched haloes as a function of the
    corresponding FOF mass, for halo constructed with different LPT
    orders, as indicated in the legend. For each order, the line in
    the middle is the median value, and the other two are the 16th and
    84th percentiles. The vertical dashed and dotted lines mark the
    mass corresponding to 100 and 500 particles.}
\end{figure}

In Fig. \ref{fig: massscatter} we show, for cleanly matched haloes,
the log of the ratio of {\pin} and FOF masses as a function of FOF
halo mass. Lines show the median and the 16th and 84th percentiles for
groups reconstructed with ZA, 2LPT and 3LPT, as specified in the
legend. It is apparent from this figure that the
higher orders give better and more unbiased masses: typical accuracy
in mass decreases form 0.18 dex for ZA to 0.13 dex for 2LPT
and 3LPT, the highest order giving no obvious
advantage. Moreover, the relative accuracy improves with mass. The
negative bias visible at the largest masses is due to the fact that we
are forcing two distributions of nearly equivalent objects to have the
same mass function; if mass is recovered with some uncertainty at the
object--to--object level, the mass functions can be equal only by
introducing a negative bias in the reconstructed masses. We will get
back to this point later.

\begin{figure}
  \centering
  \includegraphics[width=\columnwidth]{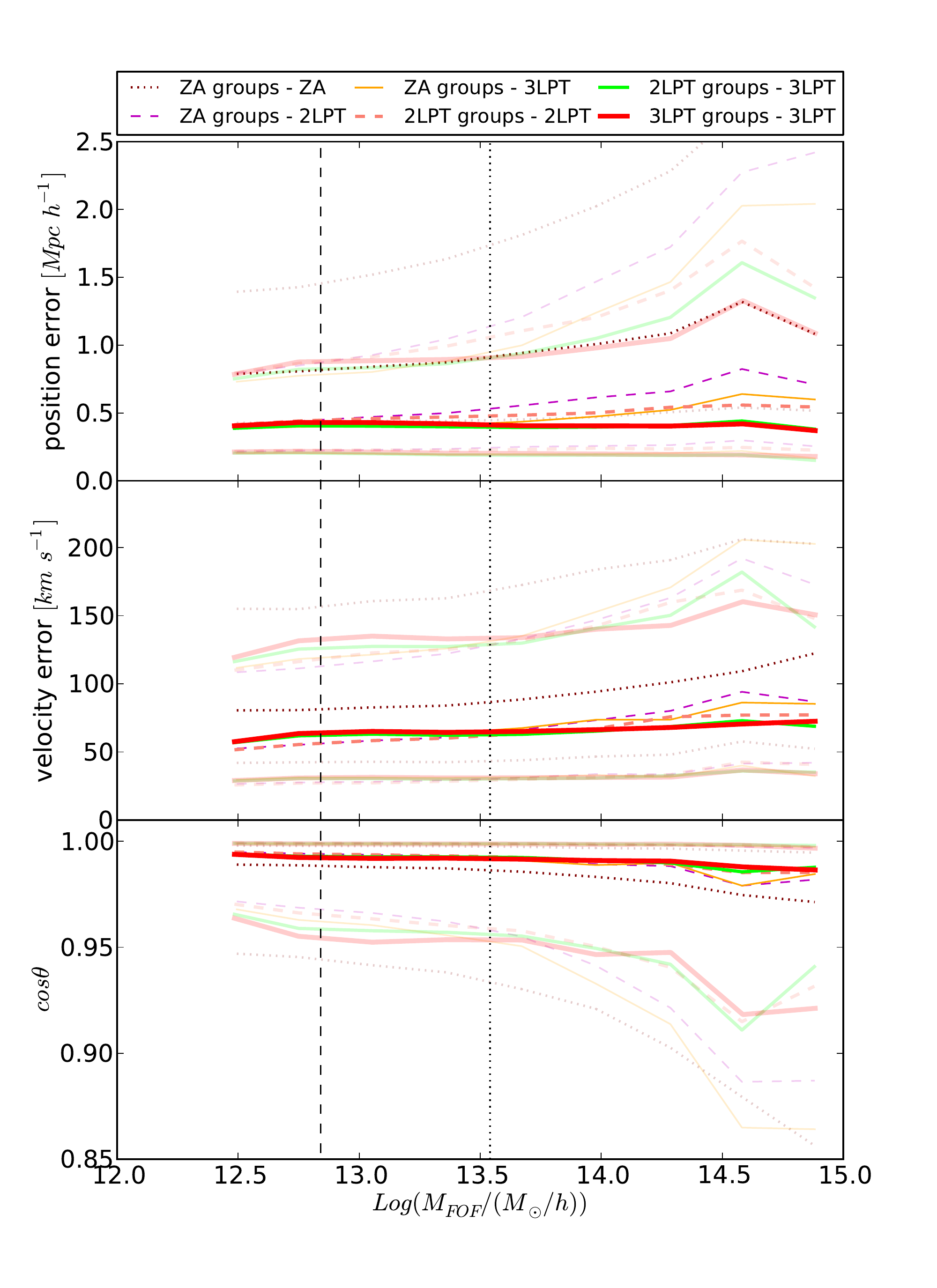}
  \caption{\label{fig: posvel} Position and velocity difference
    (\emph{top panel} and \emph{middle panel}, respectively) and
    cosine of the angle between velocity vectors (\emph{bottom panel})
    of pairs of cleanly matched haloes, for six combinations of LPT
    orders used for halo construction and halo displacement and
    velocity, as indicated in the legend.  For each configuration
    shown, there are three lines, the one in the middle being the
    median, while the other two (shown in a pale shade) being the 16th
    and 84th percentiles. The vertical dashed and dotted lines
    identify the mass corresponding to 100 and 500 particles,
    respectively.}
\end{figure}

Fig. \ref{fig: posvel} illustrates how well other halo properties are
reconstructed by {\pin}. The panel on the top shows the median (and
16th and 84th percentiles) distance between cleanly matched halo
pairs. These statistics are computed for bins of FOF halo mass, and
shown as a function of this quantity. In this figure, that involves
``Eulerian'' quantities, we show six different models with halo
construction and displacements performed with various LPT orders. The
median distances are similar for the different configurations, except
for the case of groups both constructed and displaced with ZA that
presents a systematically larger value, although groups displaced with
3LPT present a smaller median distance in most of the explored range.
The median distance in some runs has a mild positive
  dependence on halo mass, but the ratio of this distance with the
  halo Lagrangian radius is a decreasing function of halo mass, as
  discussed in Section~\ref{section:mf}.

The middle panel shows, with a very similar format, the difference in
the modulus of velocity between matched halo pairs. This quantity has a mild
dependence on mass, more massive haloes having a higher velocity
difference. In this case, 3LPT gives an improvement over 2LPT only
above $\sim10^{14} M_\odot$, while below that value 2LPT gives a
better reproduction of halo velocities.  The panel at the bottom shows
the distribution of the cosine of the angle between the velocities of
the matched pairs. In all cases velocities are very well aligned (the
median $\cos\theta$ being well below 0.95 in all cases), with the
exception of a tail of less correlated velocities. 
ZA gives on average wider angles with
a wider scatter around the median, indicating again a poorer
reconstruction of halo properties.

\section{Clustering}
\label{sect: clustering}

In this section we make use of the catalogues described in
Section~\ref{sect: catalogues}, obtained with the setup of $1024^3$
particles in a 1024 Mpc/h box, and compute their clustering
properties at $z = 0$ and $1$. In the following, we will use $k_{MAX}
= 0.5\, {\rm h\, Mpc}^{-1}$ as reference frequency beyond which we expect
non--linearities to become very important.  It is worth stressing that
the parameters of our code have already been calibrated by requiring a
good fit of the simulated mass function, so clustering of haloes is a
pure prediction of the code.

\subsection{Power spectrum}
\label{sect: Pk}
There are two main sources of disagreement in the power spectra of
{\pin} and simulations. On large scales, a difference in the linear
bias term will give a constant offset, at small scales the power
spectrum obtained with LPT displacements will drop below the simulated
one beyond some wavenumber $k$. We decided to separate these two
sources of disagreement, both to estimate in a more consistent way the
$k$--value at which the power spectrum $P(k)$ drops below a certain
level, and to stress that bias is a prediction of our code \citep[see,
  e.g.,][]{paranjape2013}.

\begin{figure*}
  \centering
  \includegraphics[width=\textwidth]{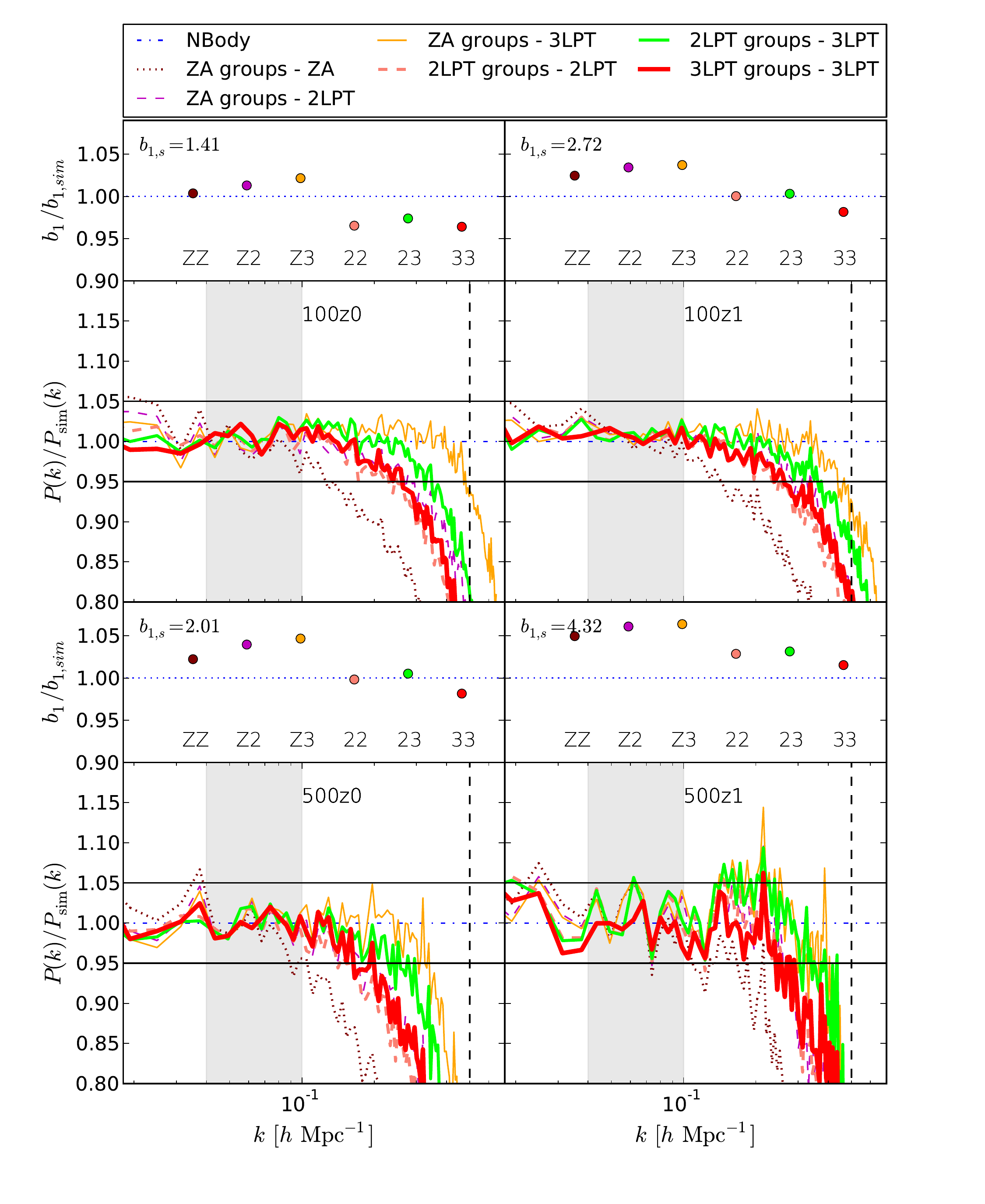}
  \caption{\label{fig: Pk} We show, for $z=0$ (left panels) and $z=1$
    (right panels) and for the mass cuts of 100 (upper panels) and 500
    particles (lower panels), the linear bias $b_1$, divided by the
    N--body's one, and the power spectrum $P(k)$ in real space,
    divided by the one of FOF catalogue $P_{sim}$ and normalized to
    unity at $k \leq 0.1 {\rm h\, Mpc}^{-1}$.  The two black
    horizontal solid lines in the power spectrum panels locate the
    $10$ per cent accuracy region. The vertical dashed line locates
    $k_{MAX}$, while the shaded area locates the region of the BAO
    peak. In the bias panels we show the bias $b_1$ normalized by the
    one of the FOF catalogue from the simulation, the value of which
    is reported in each bias panel. The dotted blue horizontal line in
    the bias panels locates the value 1.  }
\end{figure*}

In Figure~\ref{fig: Pk} we show the accuracy with which the power
spectrum $P_{\rm sim}(k)$ of a FOF halo catalogue, in real space, is
recovered by various versions of {\pin} that use increasing LPT orders
for constructing and displacing haloes.  The four groups of panels
show results for the two mass cuts of 100 (upper panels) and 500
particles (lower panels), at $z = 0$ (left panels) and $z=1$ (right
panels). Each group of panels show in the upper stripe the recovery of
the normalization of the power spectrum, quantified in this case as a
linear bias term $b_1$, that is obtained as the square root of the
average value of $P(k)$ divided by the matter power spectrum $P_m(k)$
of the simulation, computed over $k \leq 0.1 \,{\rm h\, Mpc}^{-1}$:

\begin{equation}
  P(k) = b_1^2\,P_m(k).
\end{equation}

\noindent
In the lower panels we show the ratio $P(k)/P_{\rm sim}(k) \times
(b_{1,sim}/b_1)^2$, i.e. the residuals of the {\pin} power spectrum with
respect to the N-body one, normalized to unity for $k \leq 0.1 
{\rm h\,Mpc}^{-1}$. 
We will quantify the level of agreement between N-body and {\pin}
power spectra as the wavenumber $k_{10\%}$ at which the normalized
residuals go beyond a $[0.9,1.1]$ interval.  We report 
in Table \ref{tab: k10pc} the $k_{10\%}$ values for all the cases
analysed below.

 Considering haloes constructed with ZA and displaced with
  increasing LPT order (``ZZ'', ``Z2'', ``Z3''), we notice that, in
  terms of $k_{10\%}$, higher orders give a better recovery of the
  halo $P(k)$ at all redshifts and for all mass cuts: figures raise
  from $\sim0.1-0.2$ to $\gtrsim0.4-0.5 \, {\rm h\, Mpc}^{-1}$, thus
  improving by at least a factor of two. The figure of $k_{10\%}=0.1
  {\rm h\, Mpc}^{-1}$ is consistent with the results of M13, while
  higher orders give negligible loss of power at the BAO scale.
  Regarding the bias, a clear trend of increasing $b_1$ with LPT order
  is visible; this time the ``Z3'' configuration is found to give the
  largest overestimate of $b_1$. 

 A different trend is found when 3LPT is used to displace
  the halos and the order for group reconstruction is varied (``Z3'',
  ``23'', ``33''). At $z=0$ $k_{10\%}$ decreases by a factor of two at
  $z=0$, while the effect is slightly less evident at $z=1$. Bias
  $b_1$ decreases with increasing LPT order, correcting for the
  overestimate given by the ``Z3'' configuration. Overall, the
  combination ``Z3'' of ZA groups and 3LPT displacements is the one
  that gives the best result in terms of $k_{10\%}$, but the
  combination ``23'' gives the best combination of power loss and
  bias. For this configuration $k_{10\%}$ is typically greater than
  $0.3\ {\rm h\, Mpc}^{-1}$, while bias $b_1$ is recovered at least to
  within 4 per cent in the worst case.

\begin{figure*}
  \centering
  \includegraphics[width=\textwidth]{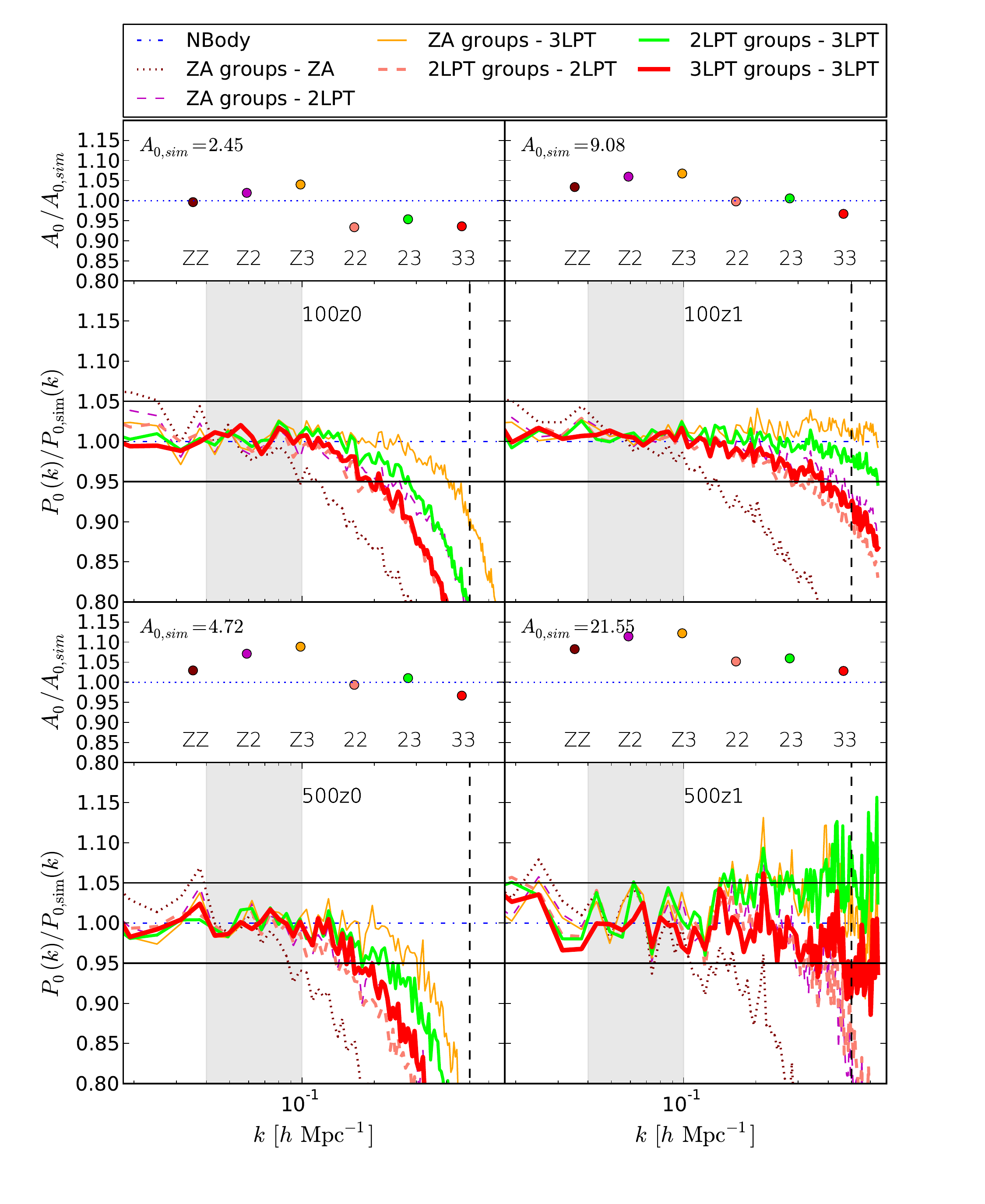}
  \caption{\label{fig: P0} Recovery of the monopole $P_0(k)$ of the 2D
    power spectrum in redshift space. Panels and symbols are
    as in Figure~\ref{fig: Pk}. Upper rows give the measured value
    of the quantity $A_0$ as defined in equation~\ref{eq:A0def}.
    }
\end{figure*}

\begin{figure*}
  \centering
  \includegraphics[width=\textwidth]{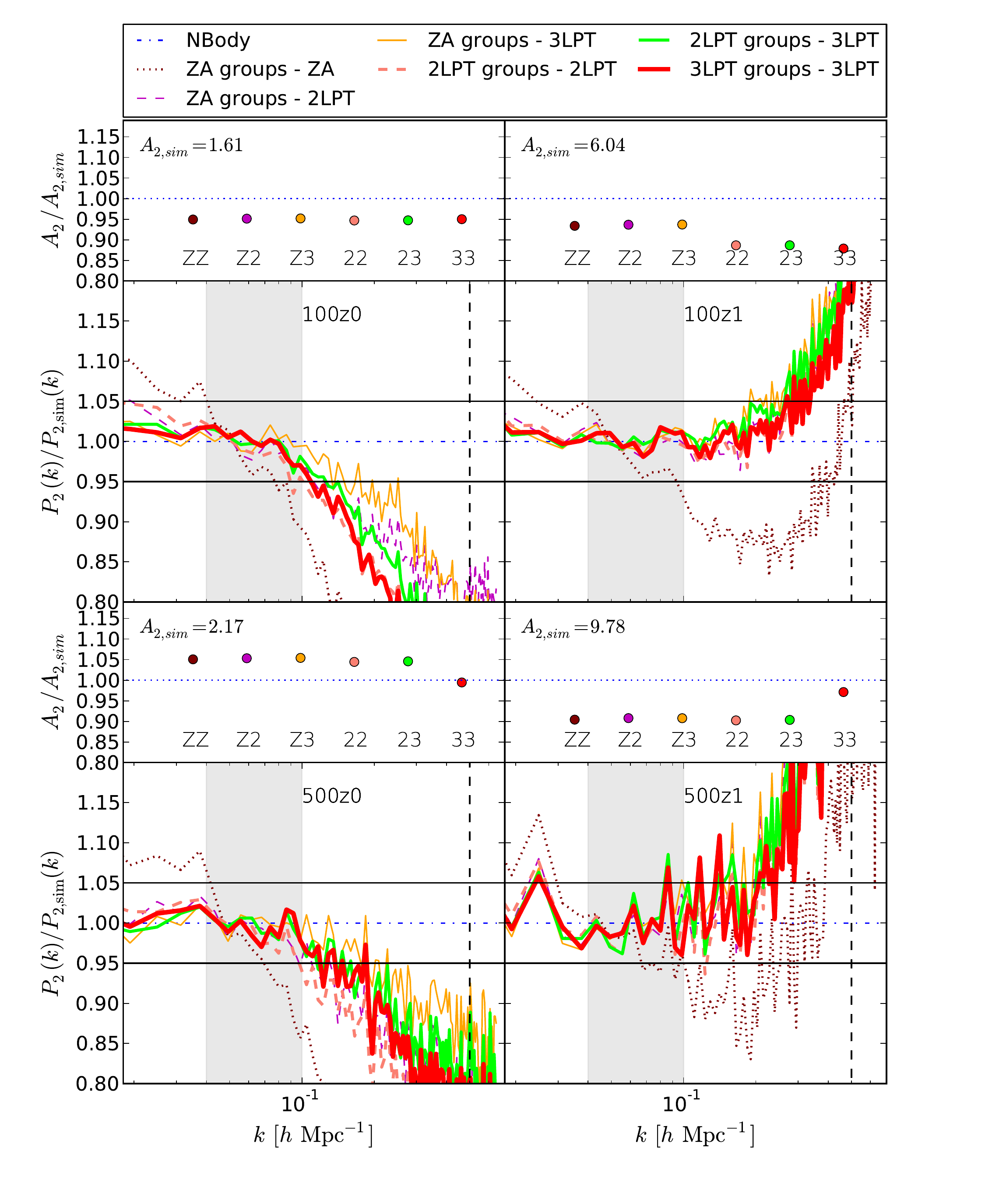}
  \caption{\label{fig: P2}  Recovery of the quadrupole $P_2(k)$ of the 2D
    power spectrum in redshift space. Panels and symbols are
    as in Figure~\ref{fig: Pk}. Upper rows give the measured value
    of the quantity $A_2$ as defined in equation~\ref{eq:A2def}.}
\end{figure*}

Figure~\ref{fig: P0} shows the monopole $P_0(k)$ of the 2D power
spectrum in redshift space. For this calculation we used the three
axes as three lines of sight, and averaged the results over the three
orientations.  The monopole is related to the matter power spectrum
via the following relation:
\begin{equation}
  P_0(k) = A_0\, P_m(k),
\label{eq:A0def}\end{equation}
where the coefficient $A_0$ in linear theory is equal to:
\begin{equation}
A_0 = \left(1+\frac{2}{3}\beta+\frac{1}{5}\beta^2\right)b_1^2,
\end{equation}
where $\beta=f(\Omega_m)/b_1$.  The results are shown in the same
format as in Figure~\ref{fig: Pk}. While the relative merits of the
combinations of LPT orders are the same as in the real space $P(k)$,
the ability to recover the mildly non-linear part of the power
spectrum at $z=1$ is significantly better, with the best methods
giving $k_{10\%}$ well in excess of $0.5 {\rm h\, Mpc}^{-1}$.  This
advantage seems to be lost at $z=0$. Again, the combinations that
  perform better are those with groups displaced by 3LPT. The
  combination ``Z3'' is the best choice in terms of $k_{10\%}$, 
  while ``23'' gives the best compromise between $k_{10\%}$ and $A_0$.
  When
  considering the high mass cut at $z=1$ the results are very noisy
  because of the poor statistics, but indicate that the ``33''
  combination is the one that performs better. The accuracy in the
recovery of the normalization $A_0$ is similar to that of $b_1$ once
one takes into account that the coefficient depends on the square of
bias. The best methods give agreement at the $5-10$ per cent level.

Fig. \ref{fig: P2} shows the quadrupole $P_2(k)$ of the 2D power
spectrum in redshift space, computed again as an average over the
three axis orientations.  The quadrupole is related to the matter
power spectrum via the following relation:
\begin{equation}
  P_2(k) = A_2\, P_m(k),
\label{eq:A2def}\end{equation}
where the coefficient $A_2$ in linear theory is equal to:
\begin{equation}
A_2 = \left(\frac{4}{3}\beta+\frac{4}{7}\beta^2\right)b_1^2.
\end{equation}

In this case {\pin} catalogues show a strong power loss at $z=0$. At
$z=1$ the tension is alleviated, the quadrupole appears compatible
with the N--body's one at low $k$, showing an excess of power for
higher wavenumbers. For this specific observable, all the
configurations, except ``ZZ'', give comparable results, possibly due
to the higher noise level that hides the fine details. In this case
the ``ZZ'' configuration performs poorly even at the BAO scale,
showing again the advantage of going to higher orders.

In linear regime, the growth rate $f$ is related to $\beta$ and to the
bias $b_1$ via the following relation: $f = \beta \cdot b_1$. If we
multiply the mean value of $b_1$ within $k = 0.1 {\rm h\, Mpc}^{-1}$ with
$\beta$, obtained by solving the ratio given by eq. \ref{eq:A0def}
divided by eq. \ref{eq:A2def} and averaged in the same range of $b_1$,
we obtain values of the growth rate that are consistent with the
actual theoretical value $f \simeq \Omega_m^{0.6}$.

\begin{figure}
  \centering
  \includegraphics[width=\columnwidth]{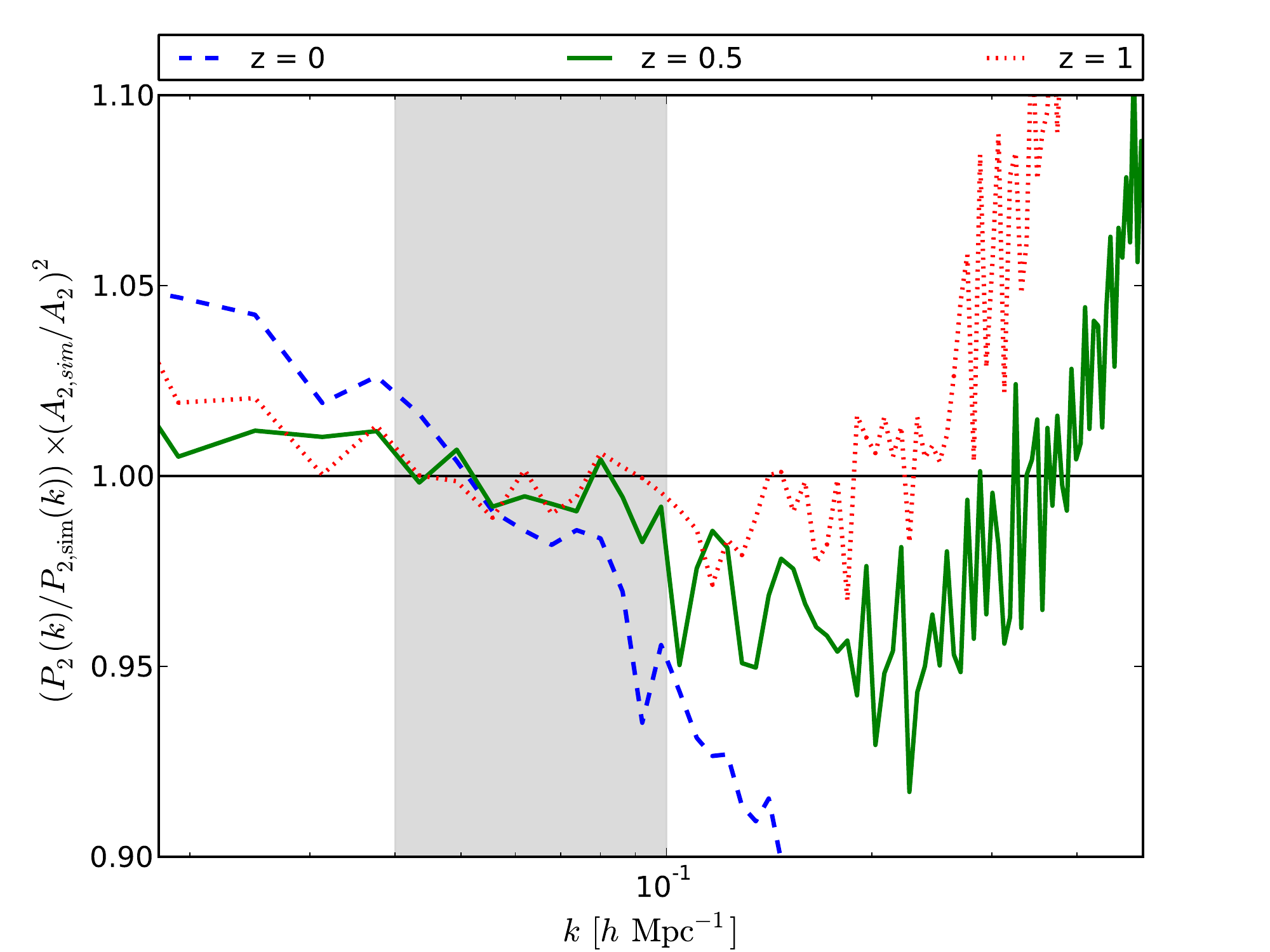}
  \caption{\label{fig: nIFTy Comparison} Recovery of the quadrupole
    for the configuration ``22'' for the catalogue ``100'' at three
    different redshifts, as indicated in the legend. The shaded area
    locates the region of the BAO peak. }
\end{figure}

To further investigate the degradation of the agreement in the
quadrupole term at low redshift, in Figure \ref{fig: nIFTy Comparison}
we show the recovery of the quadrupole for the configuration ``22''
and only for the catalogue ``100'' at three different redshifts,
namely 0, 0.5 and 1. This configuration allows a direct comparison
with what was found in \cite{chuang2015}, as in that paper the
redshift was $0.56$, the number density $3.5 \times 10^{-4}
{\rm h^3Mpc}^{-3}$, and {\pin} was run with 2LPT both for halo
construction and displacement (the number density of our ``100''
catalogue at $z=0.5$ is $3.85\cdot 10^{-4}$). The behaviour of the
quadrupole at $z=0.5$ is compatible with that found in
\cite{chuang2015}, with a loss of power that reaches a minimum of
$\sim 8$ per cent at $k \simeq 0.2 {\rm h\, Mpc}^{-1}$, and a successive gain of
power that allows the quadrupole to reach the N--body's value at $k
\simeq 0.4 {\rm h\, Mpc}^{-1}$. What is evident from Figure \ref{fig: nIFTy
  Comparison} is the evolution of the quadrupole with redshift. For
redshift above 0.5, that is very relevant for future surveys, the
quadrupole is within $10$ per cent accuracy for $k \lesssim 0.3 {\rm h\,
Mpc}^{-1}$.

\begin{table}
\centering
\begin{tabular}{lllll}
\hline
 & 100 z0 & 100 z1 & 500 z0 & 500 z1\\
\hline \hline 
\multicolumn{5}{l}{Power spectrum real space}\\
\hline 
ZA groups - ZA & 0.21 & 0.23 & 0.15 & 0.21\\
ZA groups - 2LPT & 0.39 & 0.38 & 0.24 & 0.29\\
ZA groups - 3LPT & 0.55 & 0.54 & 0.38 & 0.42\\
2LPT groups - 2LPT & 0.31 & 0.36 & 0.22 & 0.33\\
2LPT groups - 3LPT & 0.43 & 0.47 & 0.34 & 0.44\\
3LPT groups - 3LPT & 0.34 & 0.37 & 0.25 & 0.37\\

\hline
\multicolumn{5}{l}{Monopole redshift space}\\
\hline 
ZA groups - ZA & 0.16 & 0.21 & 0.13 & 0.21\\
ZA groups - 2LPT & 0.34 & 0.61 & 0.23 & 0.45\\
ZA groups - 3LPT & 0.50 & 0.64 & 0.38 & 0.64\\
2LPT groups - 2LPT & 0.28 & 0.51 & 0.20 & 0.49\\
2LPT groups - 3LPT & 0.35 & 0.64 & 0.31 & 0.64\\
3LPT groups - 3LPT & 0.28 & 0.58 & 0.23 & 0.64\\

\hline 
\multicolumn{5}{l}{Quadrupole redshift space}\\
\hline 
ZA groups - ZA & 0.09 & 0.54 & 0.09 & 0.63\\
ZA groups - 2LPT & 0.25 & 0.36 & 0.18 & 0.29\\
ZA groups - 3LPT & 0.30 & 0.36 & 0.58 & 0.29\\
2LPT groups - 2LPT & 0.14 & 0.37 & 0.18 & 0.29\\
2LPT groups - 3LPT & 0.17 & 0.35 & 0.28 & 0.25\\
3LPT groups - 3LPT & 0.15 & 0.45 & 0.21 & 0.29\\

\hline 
\end{tabular}
\caption{\label{tab: k10pc} $k_{10\%}$ values, in units of
  $\rm h\,Mpc^{-1}$, beyond which the normalized residuals of the {\pin}
  power spectra with respect to the N-body ones differ by more than
  $10$ per cent from unity.}
\end{table}

Given these results, in the following clustering analysis we focus our
attention only on the configurations that perform better in the power
spectrum analysis, that is haloes constructed with ZA and displaced
with 3LPT (``Z3'') and haloes constructed with 2LPT and displaced with
3LPT (``23'').  We will also consider the ``22'' configuration, that
has significant smaller memory requirements, and, for comparison with
previous work, the ``ZZ'' configuration of the previous version of
{\pin}, with haloes constructed and displaced with ZA.

\subsection{Bispectrum}

The distribution of DM haloes is characterised by a high level of
non-Gaussianity resulting from both the nonlinear evolution of matter
perturbation as from the nonlinear and non-local properties of halo
bias \citep[see e.g.][]{chan2012}. In particular such non-Gaussianity
is responsible for relevant contributions to the covariance of the
galaxy power spectrum \citep[see, e.g.][]{hamilton2006,deputter2012}
and it is therefore crucial that any approximate method aiming at
reproducing halo clustering can properly account for it.

Non-Gaussianity is simply defined by the non-vanishing of any (or all)
connected correlation function of order higher than the two point
function \citep[for a generic introduction
  see][]{bernardeau2002}. Here we simply consider the lowest order in
the correlation function hierarchy, limiting ourselves to a comparison
of bispectrum $B(k_1,k_2,k_3)$, i.e. the 3-PCF in Fourier space. As
opposed to sampling a few subsets of triangular configurations (as
done for instance in M13), here we attempt a comparison between all
measurable configurations defined by triplets of wavenumber
$(k_1,k_2,k_3)$ up to $0.2\,{\rm h/Mpc}$. Specifically we consider
wavenumber values defined by linear bins of size 3 times the
fundamental frequency of the box.

\begin{figure}
  \centering
  \includegraphics[width=\columnwidth]{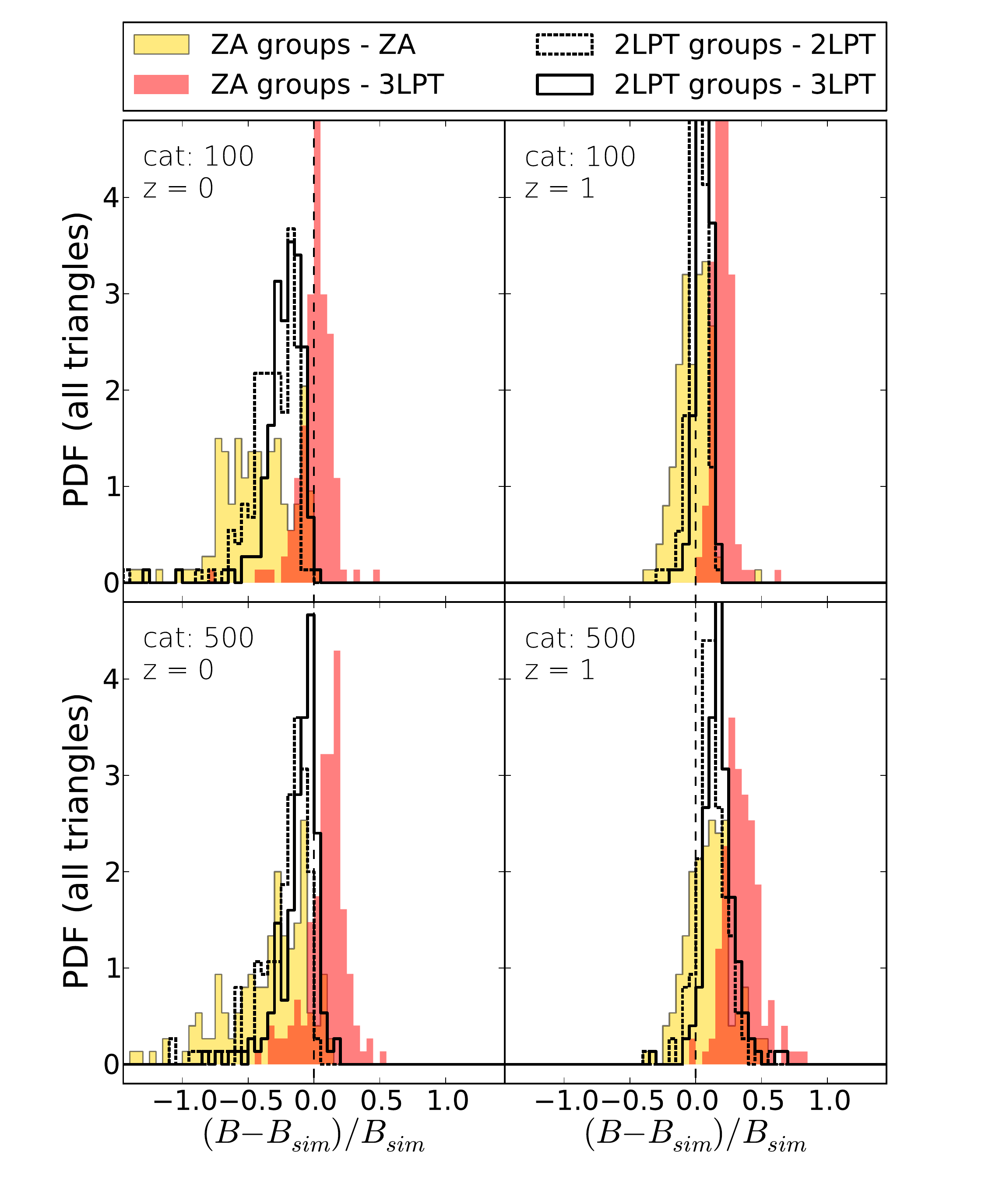}
  \includegraphics[width=\columnwidth]{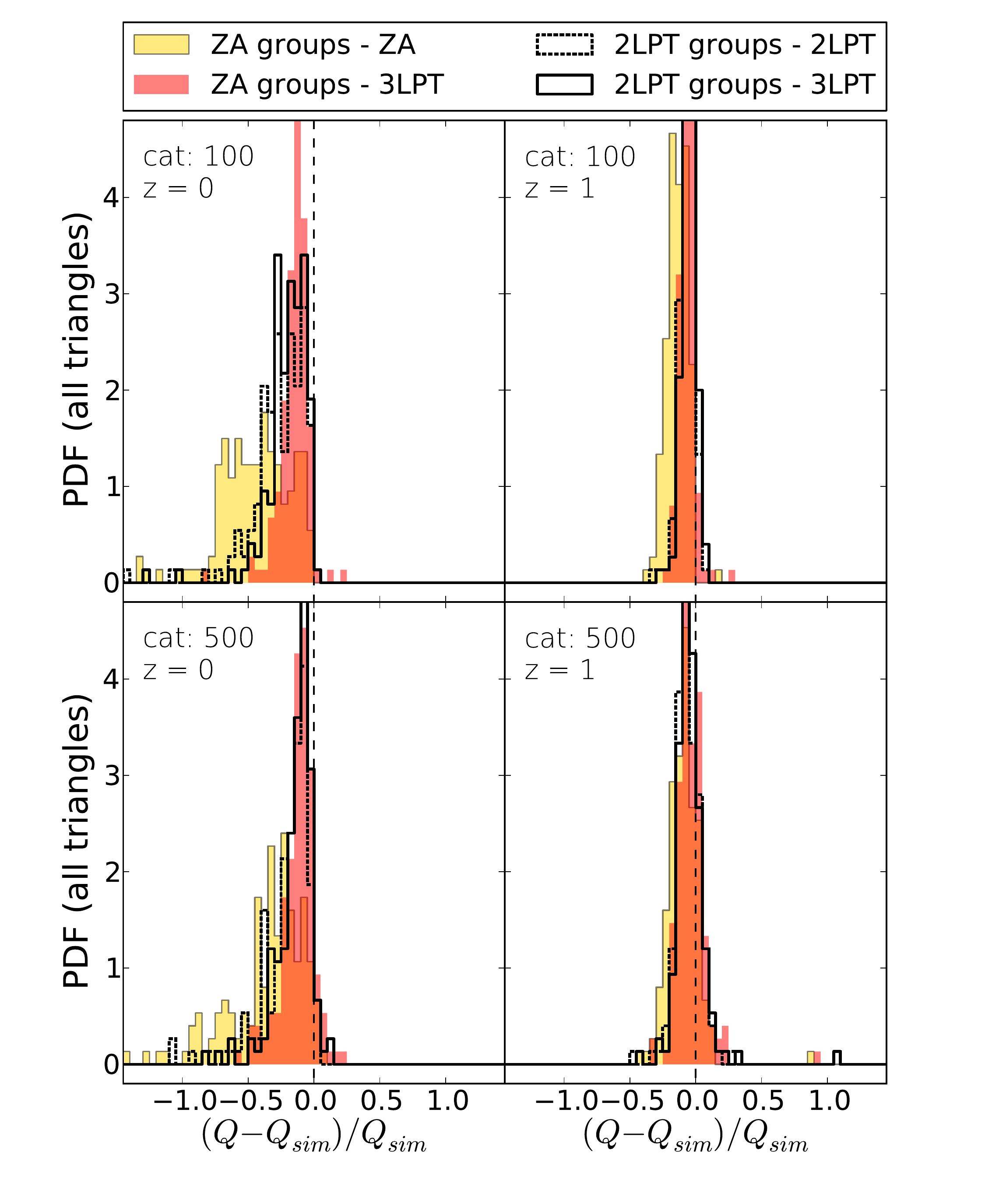}
  \caption{\label{fig: bispec} Pdf of the relative difference of the
    bispectrum (\emph{top panels}) and reduced bispectrum
    (\emph{bottom panels}) with respect to the N--body's ones for four
    configurations, as indicated in the legend. The four subpanels
    refer to $z=0$ and $1$ (\emph{left and right columns},
    respectively) and high and low mass cuts (\emph{top and bottom
      rows}, respectively), as indicated in the figures. All the
    triangles up to $k = 0.2 {\rm h\,Mpc}^{-1}$ are considered.}
\end{figure}

The top panels of Fig. \ref{fig: bispec} show the distribution
obtained from individual values of the relative difference $[
  B(k_1,k_2,k_3)-B_{sim}(k_1,k_2,k_3) ] / B_{sim}(k_1,k_2,k_3)$
evaluated for all triangular configurations $k_1,k_2,k_3$ with
$0.2\,{\rm h/Mpc}\geq k_1 \geq k_2 \geq k_3$.  It is evident that
  higher-order LPT, when compared to ZA results based on ZA
  displacements, reduces the variance of such distribution at $z=0$.
  The differences between the other redshifts and the other orders is,
  on the other hand, less evident, but in three out of four panels
  ``Z3'' is found to overestimate the relative difference, while
  ``23'' gives less biased results. 

The lower panel shows a similar comparison this time for the reduced
bispectrum $Q(k_1,k_2,k_3)$ defined as the ratio $Q(k_1,k_2,k_3) =
B(k_1,k_2,k_3) / P(k_1) P(k_2) + perm$. This quantity has the
advantage of highlighting the dependence on the triangle shape by
reducing the overall dependence on scale, particularly significant in
the penuchi result due to the lack of power at small
scales. In fact, we notice that {\pin} predictions are slightly more
accurate for the reduced bispectrum than for the bispectrum itself.

The results confirm what was found in M13, that the ability of
  the code to recover clustering extends to higher order statistics,
  and show improvements with respect to the original version based
  on ZA both for group construction and displacement.

\subsection{Phase correlations}

\begin{figure}
  \centering
  \includegraphics[width=\columnwidth]{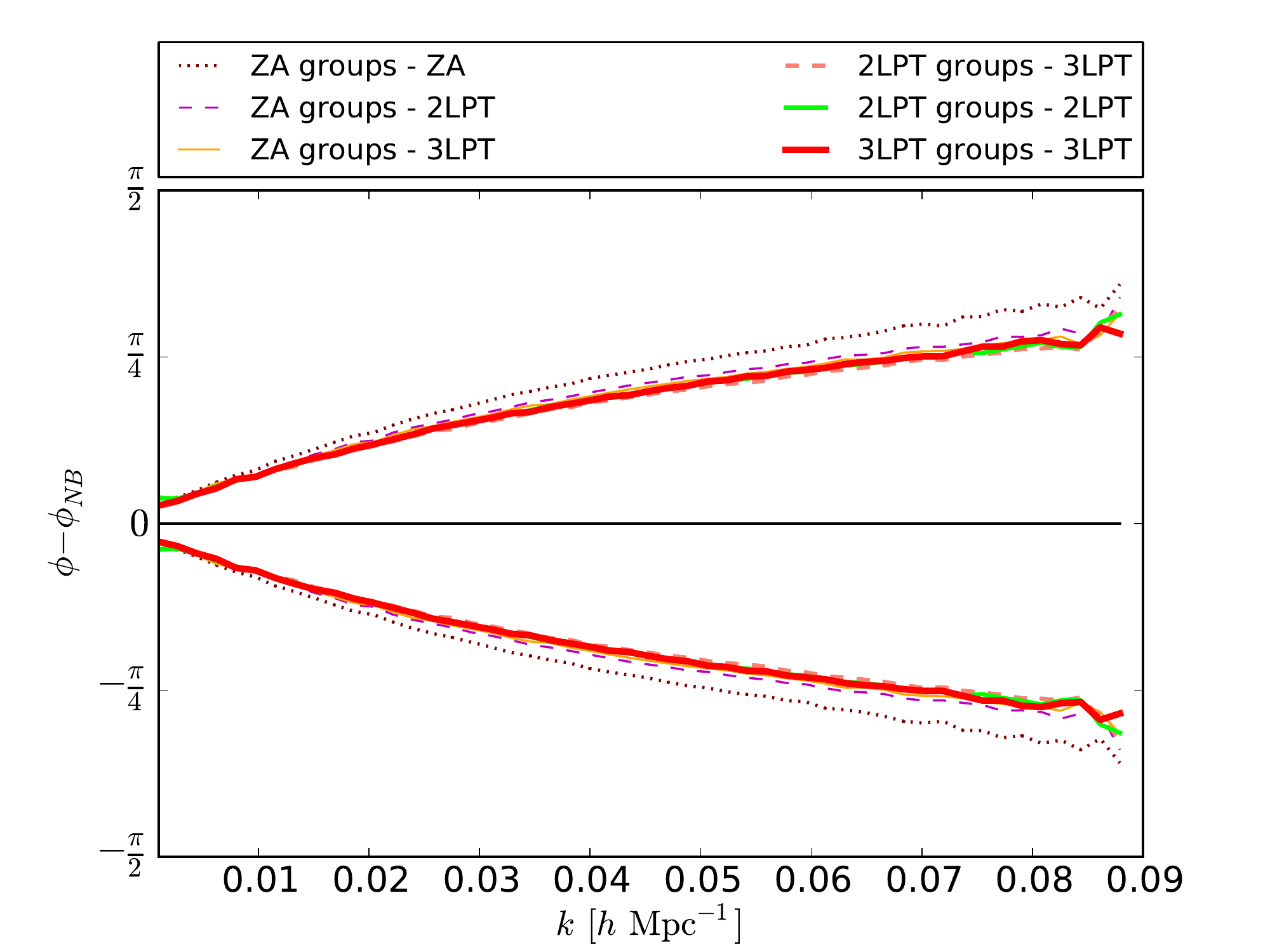}
  \caption{\label{fig: phases} Phase difference, computed on a $150^3$
    cell grid at $z=0$, between the FOF catalogue from simulation and
    the {\pin} catalogues built according to different criteria for
    the halo construction and for the halo displacements as indicated
    in the legend. The median values in each k--bin for each run are
    plotted as black solid lines that overlap one another, while the
    16 and 84 percentiles are coded as reported in the legend.}
\end{figure}

The phase difference between the halo field in {\pin} and in the
simulation could give an indication of how important stochasticity,
i.e. coupling between long and short wavelenght fluctuations, is and
how much of it is not captured by our Lagrangian schemes for the
displacements \citep{seljak2004}. In this respect the LPT order used
to construct groups plays a minor role.  Fig. \ref{fig: phases} shows
the phase difference between the FOF catalogue from simulation and the
{\pin} catalogues built according to different criteria for the halo
construction and for the halo displacements. For each catalogue we
have computed the density field of haloes by adopting a CIC (count in
cell) algorithm on a $150^3$ cell grid. After Fourier transforming the
density field, the angles between the k--vectors in the simulation and
in each {\pin} catalogue are computed.

For all the {\pin} runs, the median is compatible with 0, with
comparable scatters around it, except for the catalogue built with ZA
for both the construction and the displacements, that presents a wider
scatter. As expected, the phase difference is symmetrc around zero, a
direct consequence of the density field being a real field.

The good recovery of the phase information, or
  stochasticity, in the halo density field comes naturally from the
  good performance of the {\pin} on a object by object basis. This
  also makes {\pin} best suitable for cross correlation between
  different tracers of the dark matter density field, as opposed to
  other methods based on sampling, which are tuned to reproduce auto
  power spectra only.

\subsection{2--point correlation function}

\begin{figure}
  \centering
  \includegraphics[width=\columnwidth]{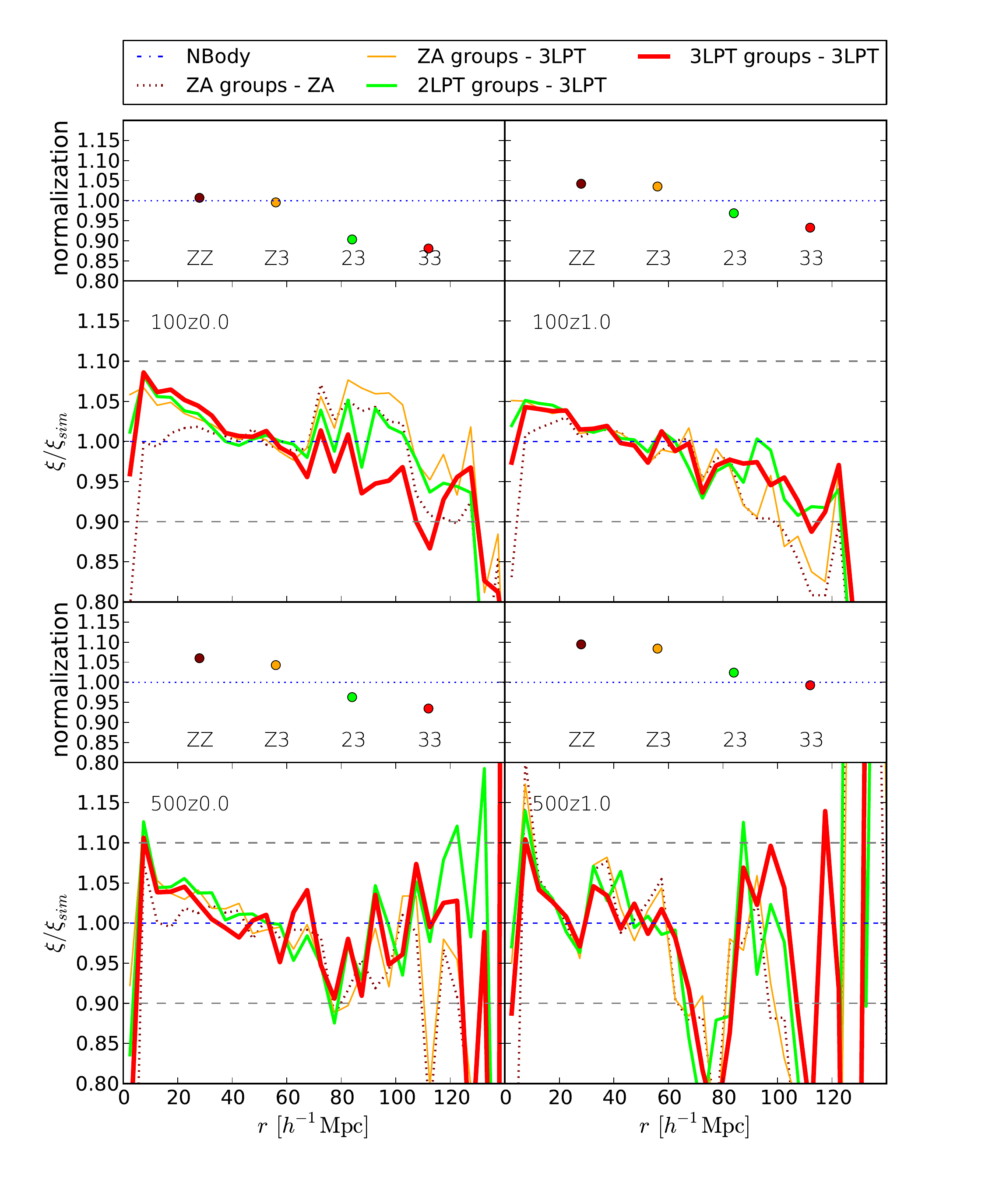}
  \caption{\label{fig: xi} 2--point correlation function in real
    space.  For the high (\emph{top panels}) and low mass cut
    catalogues (\emph{bottom panels}), at redshift 0 (\emph{left
      columns}) and 1 (\emph{right columns}), we show, in the lower
    subpanels, the ratio of the 2--point correlation function with
    respect to the N--body's one, normalized by the mean value in the
    [30-70] $h^{-1}$ Mpc interval, and the values of such
    normalization in the upper subpanels. The configurations shown are
    those indicated in the legend.}
\end{figure}

\begin{figure}
  \centering
  \includegraphics[width=\columnwidth]{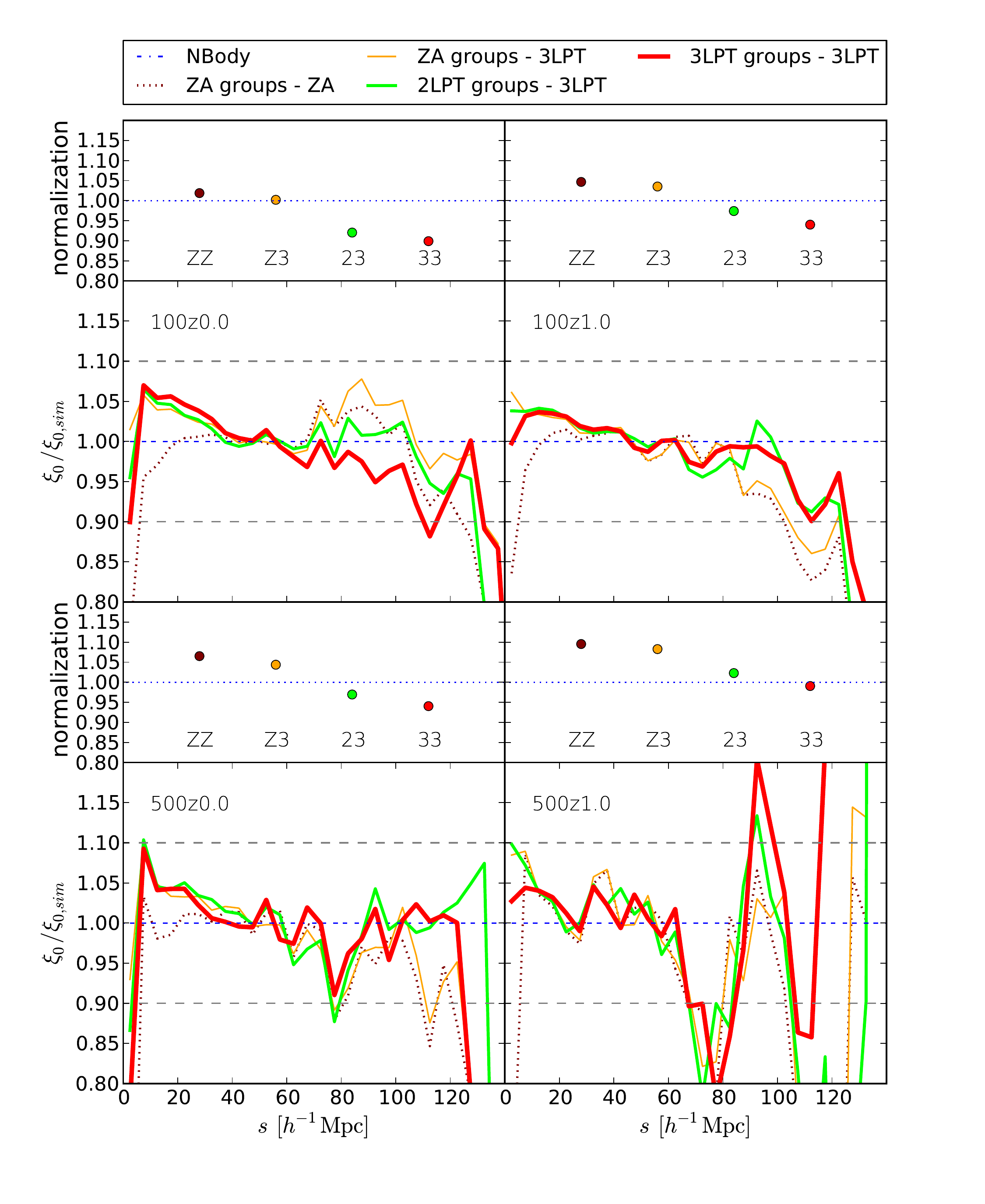}
  \caption{\label{fig: xi0} Same as \ref{fig: xi0} but for the
    monopole in redshift space.}
\end{figure}

\begin{figure}
  \centering
  \includegraphics[width=\columnwidth]{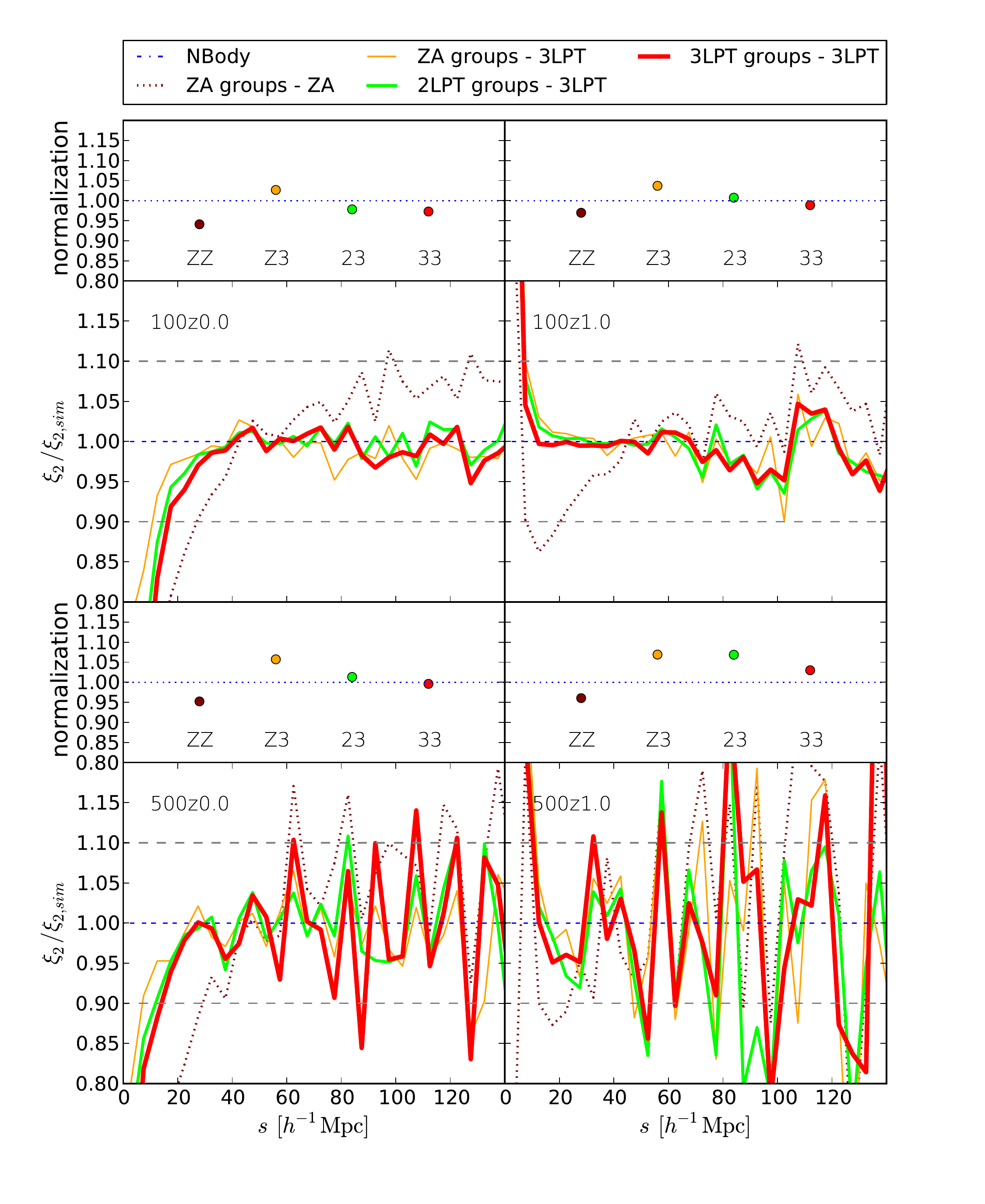}
  \caption{\label{fig: xi2} Same as \ref{fig: xi0} but for the
    quadrupole in redshift space.}
\end{figure}

Figures~\ref{fig: xi} to \ref{fig: xi2} show the 2-point correlation
function in real and redshift space (monopole and
quadrupole). Following \cite{manera2010}, here we rescale
  the correlation functions in the [30-70] $h^{-1}$ Mpc interval to
  match the large scale bias, as done in Sect. \ref{sect: Pk} for the
  power spectrum. The correlation function, both in real and in
  redshift space, is well reproduced by the different configurations,
  being within $10$ per cent of the N--body's one for most of the
  explored radial range, with the obvious exception of the
  zero-crossing region (where the correlation function reaches the
  value 0, at $~120-140 h^{-1}\rm Mpc$) and of the first radial bin,
  where the clustering is underestimated. While runs with groups
  constructed with ZA appear to have little or no bias, runs with
  groups constructed with higher LPT orders present a $5-10\%$
  bias. Besides that, differences between runs are small so that it is
  hard to decide which choice is best; the comparison in Fourier space
  gives a much clearer view of the differences.

\subsection{The origin of inaccuracy in the recovery of the linear bias term}
\label{section:bias}

{\pin} is able to recover halo masses with a good accuracy, as shown
in Figure~\ref{fig: massscatter}, where the average mass is in good
agreement with the one from simulation. However, the recovery is
subject to scatter of $\sim 0.15$ dex. At the same time, the mass
function is fit to within a 5-10 per cent accuracy.  Because most of
the recovered haloes are closely matched, an unbiased mass recovery
would lead to an overestimate of the mass function, and this justify
the little underestimate in the average mass visible in that figure,
especially at large masses (where the mass function is steeper and
then more subject to this effect).  When applying mass
  cuts, different haloes will be selected and, given the steepness of
the mass function, it is more likely that a halo is up--scattered to
higher values of mass rather than it is down--scattered to smaller
values. This will induce an underestimate in the halo bias measured
for the sample, that could justify some of the differences found in
our previous analysis.

We have quantified this effect as follows.  We have perturbed the
masses of the FoF catalogue from simulation with a log-normal
distribution with zero mean and dispersion as found in
Figure~\ref{fig: massscatter} for the ZA case. Then we have applied
{the usual mass cuts of 100 and 500 particles}.  Figure~\ref{fig: Pk
  perturbed} shows the ratio of the power spectra $P(k)$ in real space
of the perturbed catalogues and the original ones.  Clearly, this time
the residuals are not normalized to unity on large scales, and the
insets give the mean of the plotted ratios. The perturbed catalogues
present a relative (squared) bias of $\sim 3-5$ per cent at $z=0$, and
of $5-6$ per cent at $z=1$.  The error in the recovery of bias that
we have quantified above are larger than these numbers, so this effect
can justify only part of the discrepancies that we find.

\begin{figure}
  \centering
  \includegraphics[width=\columnwidth]{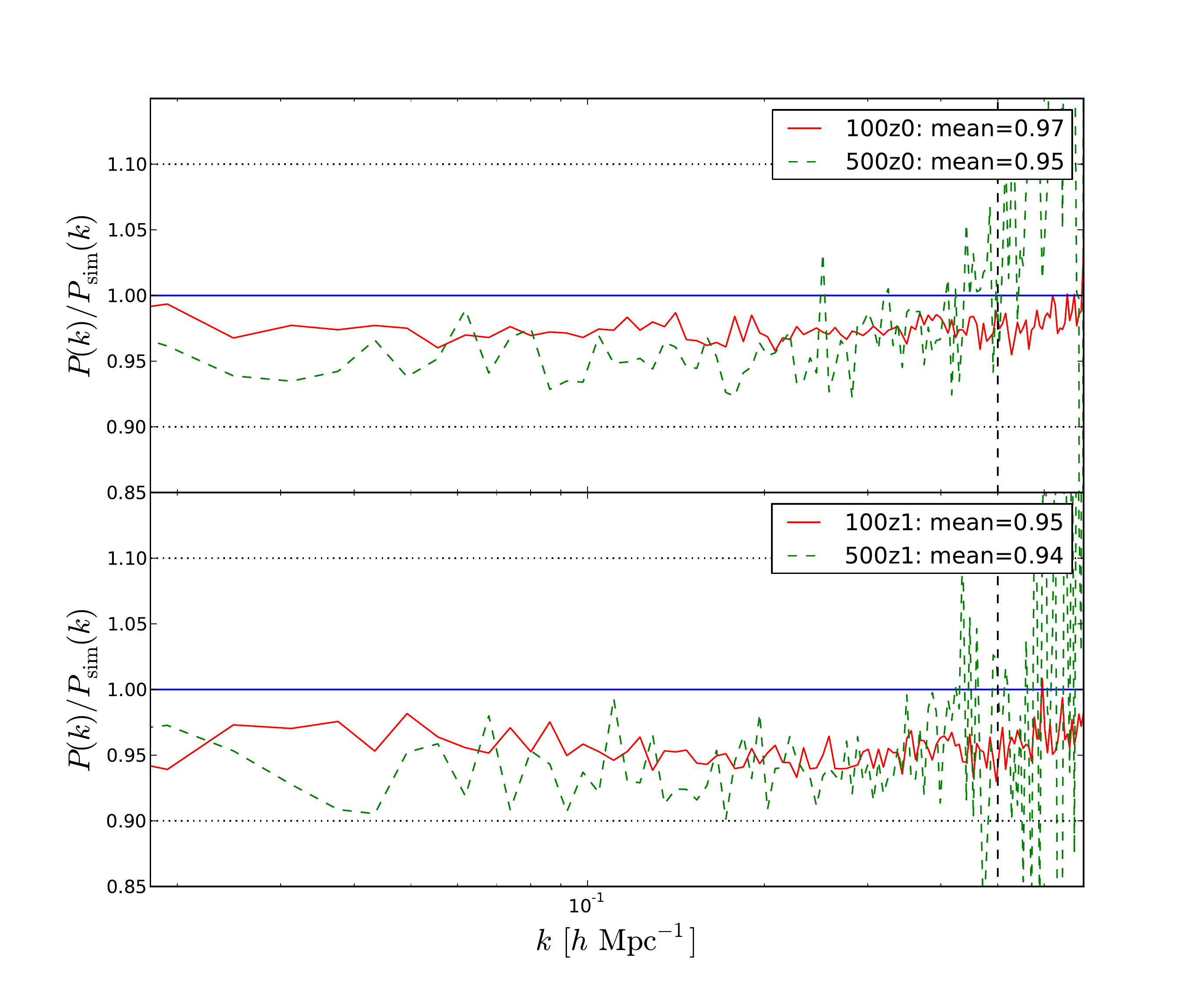}
  \caption{\label{fig: Pk perturbed} Ratio between the power spectrum
    of the perturbed FOF catalogues and the unperturbed one. Solid red
    lines refer to the high number density catalogues, while green
    dashed lines refer to the low density ones. The horizontal blue
    line locates the value 1, while the horizontal dotted ones locate
    the $10$ per cent accuracy region. The vertical dashed line
    locates $k_{MAX}$. The \emph{top panel} shows the results at
    $z=0$, while the \emph{bottom panel} at $z=1$. In the legend the
    mean values of the ratios, computed for the values below $k = 0.3
    {\rm h\,Mpc}^{-1}$, are reported.}
\end{figure}

\begin{figure}
  \centering
  \includegraphics[width=\columnwidth]{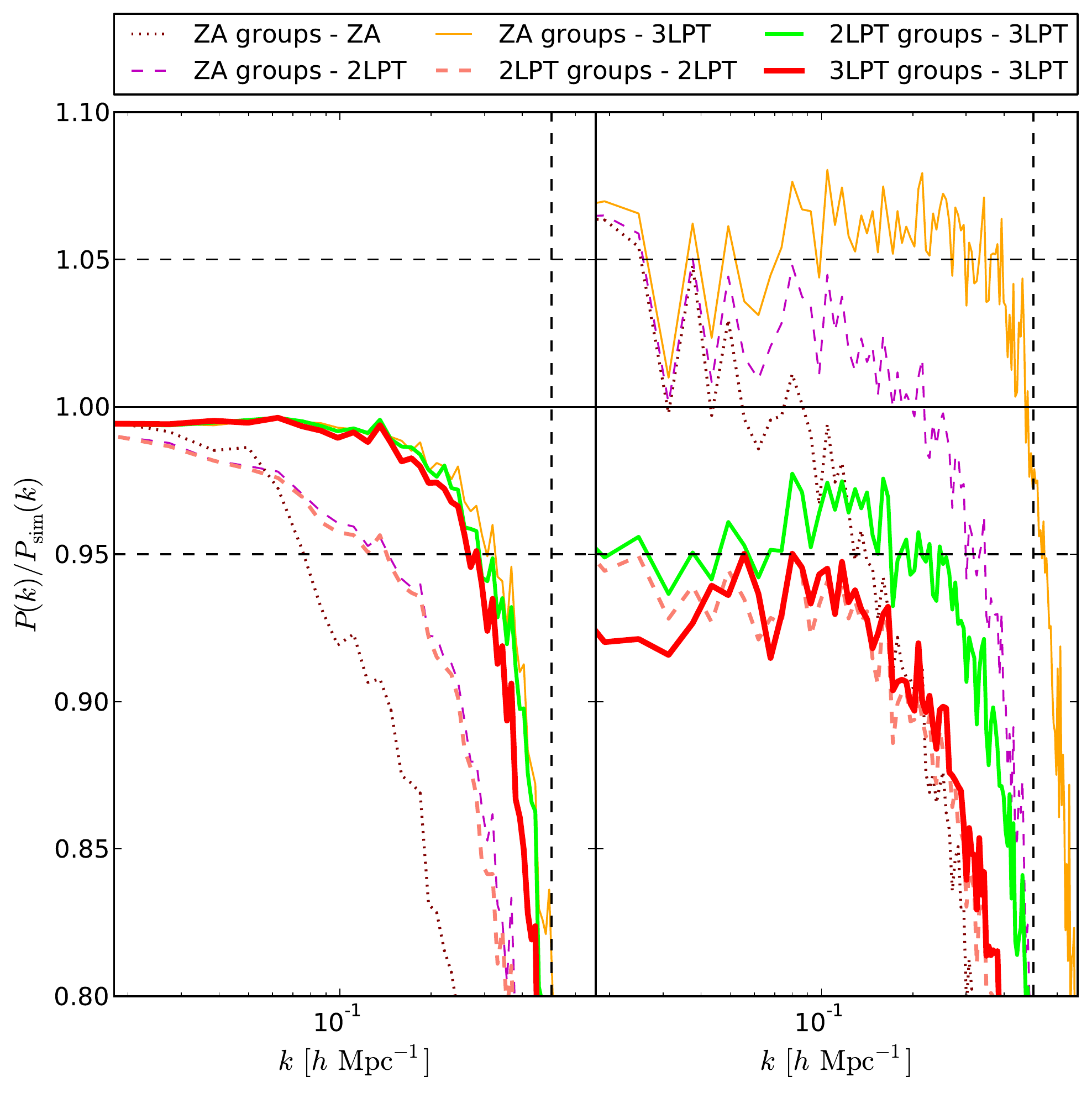}
  \caption{\label{fig: Pk Matched} Comparison between the power
    spectrum in real space at $z=0$ of the catalogue made only by
    matched haloes in the $512^3$ particle -- 512 Mpc/h side box setup
    (\emph{left panel}) and that with all haloes (\emph{right panel}),
    namely the catalogue ``100'' described in Sect. \ref{sect:
      catalogues}. Power spectra are normalized by the power spectra
    extracted from the corresponding catalogues of the N--body
    simulation. The horizontal solid line marks the value 1, while the
    horizontal dashed one located the $10$ per cent accuracy limit. The
    vertical dashed line locates $k_{MAX}$. In the legend, alongside
    the name of each configuration we report the mean value of the
    power spectrum within $k = 0.1 {\rm h\,Mpc}^{-1}$, the first value being
    relative to the matched haloes catalogues and the second one to the
    ``100'' catalogue.}
\end{figure}

We now demonstrate that the inaccuracy in the recovery of bias is
driven by mismatches in halo recovery. To this aim we use
  the setup with $512^3$ particles, taking advantage of the
  object-by-object match performed in Section~\ref{sect: object by
    object}, and we consider the mass cut corresponding to 100
  particles at $z=0$. We first restrict the catalogue to cleanly
  matched halo pairs, so that we have the same number of objects in
  the two catalogues and no difference can be ascribed to the effect
  described above.  We compute the power spectrum in real space at
$z=0$ for these sets of haloes, and we repeat the procedure for all
combinations of LPT order.  It is worth stressing that the N-body
catalogues in this case are different for each group construction, as
matching depends on how haloes are constructed.  In the left panel of
Figure~\ref{fig: Pk Matched} we show the ratios of $P(k)$ of {\pin}
and N-body catalogues; again, ratios here {\em are not} normalized to
unity at large scales.  The bias in the power spectrum of the cleanly
matched haloes is recovered to within a few per cent, especially when
higher orders are used: haloes displaced with 3LPT are within 1 per
cent of the power spectrum of the simulation within $k = 0.2 {\rm
  h\,Mpc}^{-1}$, meaning that $b_1$ is accurate by 0.5 per cent.

Conversely, the right panel of Figure~\ref{fig: Pk Matched} shows the
ratios of power spectra of {\pin} and FoF haloes (not normalized to
unity at large scales) for the whole catalogue, including mismatched halos.
The ratios range from 0.93 to 1.07, consistent with the bias values
found above, and are comparable but larger than the effect
  quantified in Figure~\ref{fig: Pk perturbed}.  The bias is therefore
  strongly affected by mismatches in halo reconstruction.  Here
  reconstruction with ZA, that is the least accurate at the
  object-by-object level, leads to a 7\% boost that is present even at
$k\sim0.4\ {\rm h\, Mpc}^{-1}$.

\section{Conclusions}
\label{section: conclusions}

We have tested the latest version of the {\pin} code, with
displacements computed up to the third order of LPT.  We have compared
the results of an N-body simulation with those obtained by running
our code on the same initial configuration, so that the agreement
extends to the object--by--object level. We have then quantified the
advantage of going to higher LPT orders, making a distinction between
halo construction and displacement at catalogue output.

The main results are the following.
\begin{itemize}
\item The mass function is used as a constraint to calibrate
  parameters that regulate accretion and merging.  This calibration is
  performed so as to reproduce the mass function of FoF haloes to
  within a few per cent, and is cosmology independent, so it is performed once
  for all. Halo clustering is entirely a prediction of the code.

\item We have compared {\pin} and simulated haloes at the
  object--by--object level. The construction of the haloes, in terms of
  number of particles, does not depend strongly on the order used for
  the halo construction, although with ZA shows a poorer level of agreement.
  with respect to the
  N--body's one. For cleanly matched objects, increasing the
  displacement order leads to halo positions and velocities closer to
  those of the corresponding halo in the N--body simulation.

\item We have computed the power spectrum in real space and the
  monopole and quadrupole of the 2D power spectrum in redshift space
  of catalogues produced with increasing LPT orders for halo
  construction and displacement, comparing them to those of the
  N--body simulation. Catalogues are constructed to have the same mass
  cuts (100 and 500 particles), and two redshifts ($z=0$ and $1$).  We
  have quantified separately the accuracy with which the normalization
  is recovered on large scales, as a linear bias term $b_1$ for real
  space and constants $A_0$ and $A_2$ for redshift space, and the
  wavenumber $k_{10\%}$ at which discrepancies in the power spectrum
  amount to $10$ per cent.

\item Higher LPT orders give significant improvements to the accuracy
  of halo clustering.  3LPT generally provides the best agreement with
  N--body when it is used to displace haloes, while lower
    LPT orders are better when used for halo construction.  Linear
  bias is typically recovered at the few per cent level when higher
  orders for the displacements are used, while $k_{10\%}$ is found in
  the range from 0.3 to 0.5, if not higher. The quadrupole is
  recovered to a similar level of accuracy at $z=1$, but suffers
  significant degradation between $z=0.5$ and $z=0$.

\item Good agreement is confirmed by an analysis of the bispectrum,
  where again higher orders give improved accuracy.  The improvement
  of higher LPT orders is confirmed by an analysis of phase
  correlations.

\item The 2--point correlation function is well reproduced within $10$
  per cent in most of the explored radial range, although
    runs with haloes constructed with higher LPT orders present a
    $5-10\%$ bias.

\item We have investigated the reason for the few per cent discrepancy
  in the recovery of linear bias. Part of it ($\sim 2-3$ per
    cent) is due to the scatter in the reconstructed halo masses,
    coupled with the selection criterion,
    and a comparable amount ($\sim 3-4$ per cent) is due to mismatches
    in halo definition between {\pin} and FoF.

\end{itemize}

These results confirm, from the one hand, the validity of the {\pin}
code in predicting the clustering of DM haloes to a few per cent level
and on scales that are not deeply affected by non-linearities. On the
other hand, it shows that higher LPT orders give a definite advantage
in this sense, and that they provide predictions that are
  accurate enough, even in the era of high-precision cosmology, for
  applications like the construction of covariance matrices. We
conclude that, though 2LPT provides already a good improvement with
respect to ZA and though 3LPT almost doubles memory requirements, the
latter can be considered as a very good choice for
  displacing the haloes. From another point of view, collapse times
are computed by solving the collapse of a homogeneous ellipsoid using
3LPT; it must be stressed that 3LPT in this context is different from
the one used to compute displacements (that contain a much higher
degree of non-locality, while ellipsoidal collapse depends entirely on
the Hessian of the peculiar potential), but consistency in the LPT
order is welcome. Moreover, 3LPT is the lowest LPT order that allows
to compute consistently in perturbation theory the four-point
correlation function, that determines the covariance of two-point
statistics, one of the most obvious applications of an approximate
code like {\pin}.

We have also considered separately the effect of 3LPT in constructing
haloes and in displacing them. The highest order provides
  noticeably advantages at displacing haloes, while for halo
  constructions lower orders perform better. Though the best results
  in term of $k_{10\%}$ are obtained with the ``Z3'' combination,
  where halos are constructed with Zel'dovich approximation and
  displaced with 3LPT, we recommend ``23'' as the best option, because
  it gives the best combination of bias and power loss at small
  scales, besides a less biased bispectrum, and because the gain in
  power of ``Z3'' is due to mismatched halos (Figure~\ref{fig: Pk
    Matched}). Indeed, the ZA gives the worst performance at the
  object-by-object level, and the worst mass function at high
  redshift. 

The loss of accuracy at higher LPT order in this case is not
  completely unexpected. Halo construction is based on an
extrapolation of LPT to the orbit crossing point, that is to the point
when the perturbation approach breaks, and then its validity degrades
with a higher level of non linearity. In this context, higher order
terms are not guaranteed to give an improvement. Conversely, halo
displacement at catalogue output implies an average over the whole
halo, that is an average over the whole region that goes into orbit
crossing. In a forthcoming paper (Munari et al., in preparation), we
will investigate how several approximate methods to displace particles
are able to reproduce the clustering of haloes, starting from the
knowledge of the particles that belong to the haloes in the
simulation. A result that we anticipate is that 3LPT is not able to
concentrate halo particles into a limited, high-density region, but it
is able to displace the haloes with very good accuracy, once positions
are averaged over the multi-stream region. This confirms that 3LPT may
not be optimal for reconstructing haloes, but it is very effective in
placing their centre of mass into the right position.

\section*{Acknowledgements}
The authors thank Volker Springel for providing us with {\sc gadget
  3}, and Giuseppe Murante for useful discussion. Data postprocessing
and storage has been done on the CINECA facility PICO, granted us
thanks to our expression of interest. Partial support from Consorzio
per la Fisica di Trieste is acknowledged. E.M. has been supported by
PRIN MIUR 2010-2011 J91J12000450001 ``The dark Universe and the cosmic
evolution of baryons: from current surveys to Euclid'' and by a
University of Trieste grant. P.M. has been supported by PRIN-INAF 2009
"Towards an Italian Network for Computational Cosmology" and by a
F.R.A. 2012 grant by the University of Trieste. S.B. has been
supported by PRIN MIUR 01278X4FL grant and by the ``InDark'' INFN
Grant. SA has been supported by a Department of Energy grant
de-sc0009946 and by NSF-AST 0908241.

\appendix
\section{Implementation and performance}
\label{Appendix: Implementation and performance}

With respect to v3.0, presented in M13,
{\pin} v4.0 has been completely re-written in C, and re-designed to
achieve an optimal use of memory.  These are the main improvements.

(1) LPT displacements are computed as follows \citep[e.g.][]{catelan1995}:

\begin{eqnarray}
  \lefteqn{\vec x(t) = \vec q 
  + D(t) \vec S^{(1)}(\vec q) 
  + D_2(t) \vec S^{(2)}(\vec q) +} \\ 
  &+ D_{3a}(t) \vec S^{(3a)}(\vec q)  
  + D_{3b}(t) \vec S^{(3b)}(\vec q) \nonumber
\label{eq:lptdispl}
\end{eqnarray}

\noindent
where $\vec q$ and $\vec x$ are the Lagrangian and Eulerian
coordinates, respectively, and the $D_{i}$ and $\vec S^{i}$ terms are
the factorized time and space parts of the LPT terms.  The $S^{i}$
terms can be expressed as gradients of a potential, the 3c rotational
term is typically very small and is neglected here. The equations for
these terms are reported in \cite{catelan1995}.  To be able to compute
the displacement of a particle at a generic time $t$, it is then
necessary to store four vectors for each particle.  Consistently with
the original algorithms, the spatial parts are computed at the same
smoothing scale at which the particle is predicted to collapse; we
have verified that results are very similar if this assumption is
dropped and displacements are computed on the unsmoothed density
field.  As discussed in the main text, different orders for
displacements can be applied at halo construction and catalogue
output.  Memory requirements amount to $80$ bytes per particle for ZA,
$145$ for 2LPT and $250$ for 3LPT, with slightly higher figures to
allow for the memory overhead described in point (2) below.  As for
the computing time, higher orders require more memory and will then be
distributed on a higher number of MPI tasks; we have noticed that the
computing time increases less rapidly than memory requirement, so
higher order do typically require slightly less elapsed time to be
performed.

(2) The second part of the code, the fragmentation of collapsed medium
into haloes, is parallelized by dividing the simulation volume into
sub-volumes, and fragmenting the sub-volumes without any further
communication. To avoid border effects, a boundary layer must be taken
into account in order to correctly construct haloes near the border.
The width of this layer is scaled according to the Lagrangian size of
the largest object expected to be present in the simulated
volume. This size can be rather large at $z=0$, amounting to $\sim30$
Mpc. This means that relatively small volumes at high resolution might
require a significant memory overhead due to these boundary layers,
especially if the computation is distributed over many cores.  To
limit this problem, fragmentation is performed by dividing the volume
in a few slabs and distributing those over tasks, thus performing the
fragmentation in steps.  This further sub-division of the simulation
volume decreases the memory requirement for each sub-volume while
increasing the overhead (being the size of the boundary layer fixed),
but it gives in general an advantage, allowing in most cases to
perform the fragmentation.

 (3) Displacements are now computed on the unsmoothed density
  field. In previous versions, particle displacements were computed at
  the same smoothing radius at which the particle collapses. But in
  fact halo displacements are obtained by averaging over halo
  particles, and this is in itself a sort of smoothing, that would
  cumulate with the previous one. Moreover, computing LPT
  displacements several times for several smoothing radii gives a
  significant overhead, that can be avoided. As a matter of fact,
  differences between the two schemes (computing displacements at each
  smoothing or only for the unsmoothed field) are minor.

(4) An algorithm has been devised to output a catalogue of haloes on
the past light cone. As a starting point, the position of a halo can
be easily computed at any time as long as its mass does not
change. Then, each time halo mass is updated, a check is performed on
whether the halo has crossed the light cone since last time it was
checked. This check is repeated for a set of periodic replications of
the box that cover the required survey geometry. This algorithm
therefore provides a continuous reconstruction of the past light cone,
without resorting to a finite number of snapshots.  The reconstruction
of the past light cone will be tested in a future paper.

(5) With respect to the naive implementation of M13, the
redistribution of memory from the plane-based scheme of the first part
of the code to the sub-volume-based scheme of the second part is done
using a hypecubic communication scheme. Being this very efficient,
communications are subdominant in any configuration.

\begin{figure*}
  \centering{
  \includegraphics[width=0.4\textwidth,angle=-90]{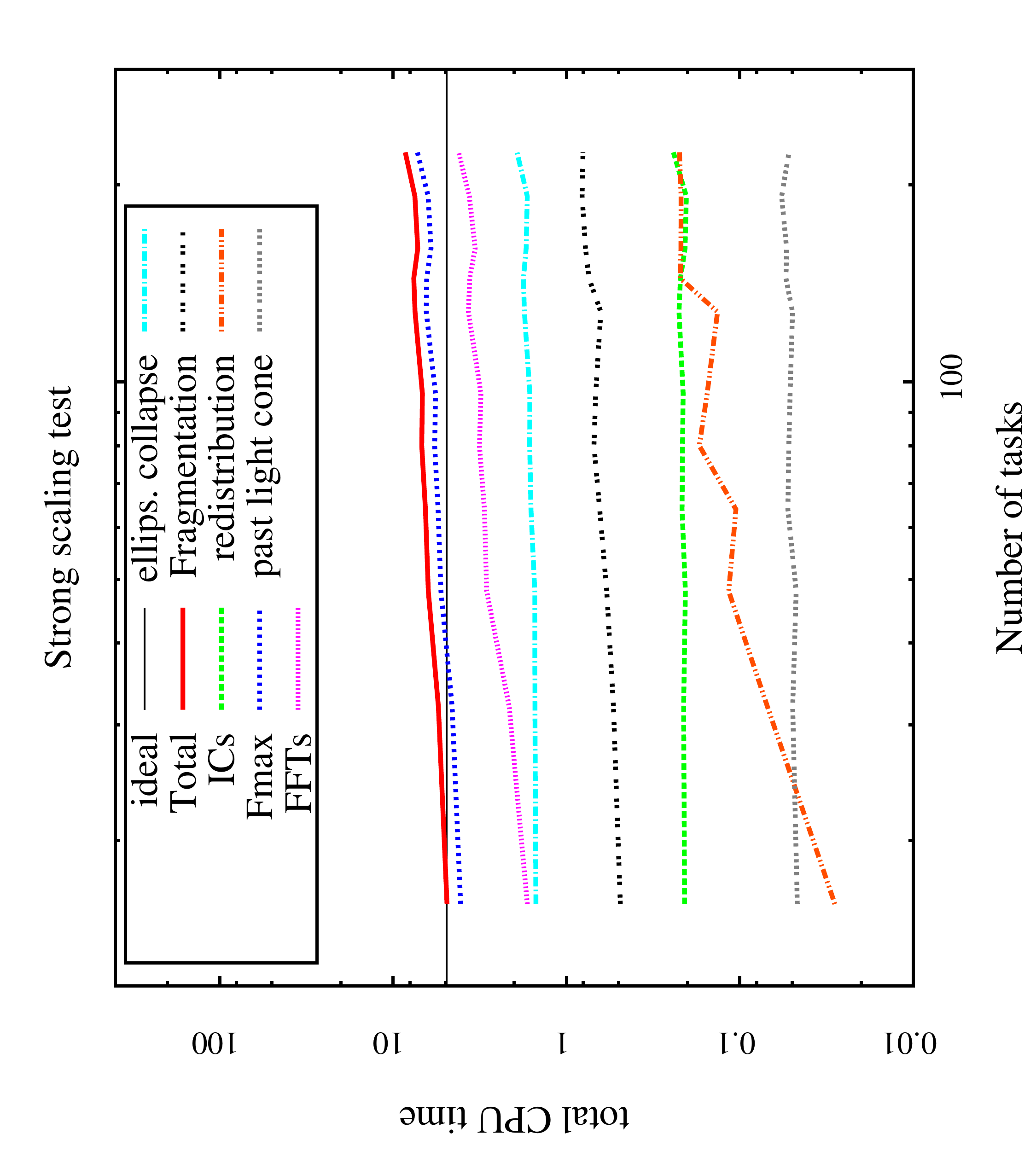}
  \includegraphics[width=0.4\textwidth,angle=-90]{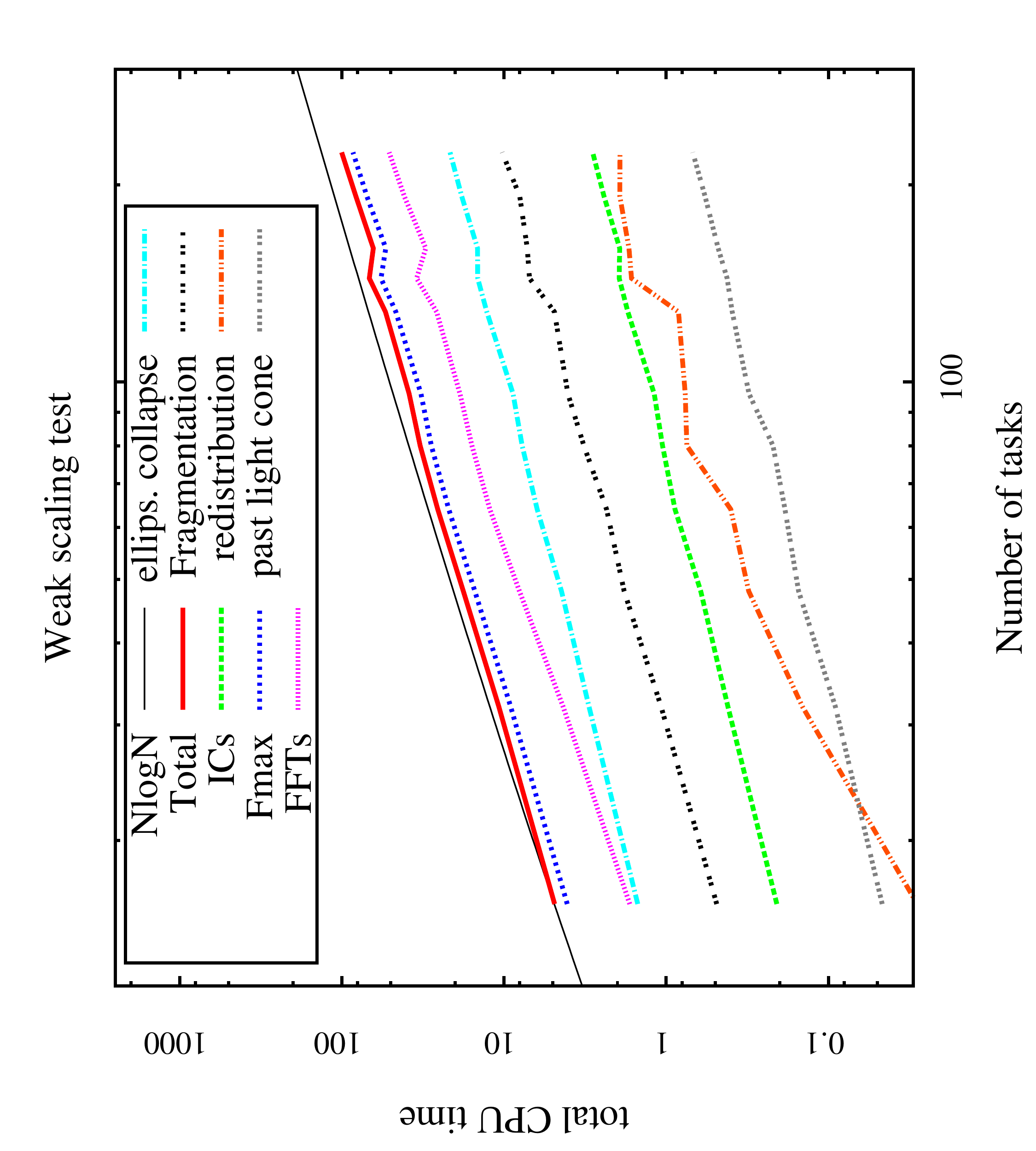}}
  \caption{\label{fig:scalingtest}Strong (left) and weak (right)
    scaling tests for the code, in the configuration specified in the
    main text.  The various lines give the computing time required by
    various parts of the code, as specified in the legend: initial
    conditions (the initial density field), ``Fmax'' (the computation
    of inverse collapse times for all smoothing radii), FFTs, the
    computation of ellipsoidal collapse, fragmentation (or halo
    construction), redistribution of memory from planes to
    sub-volumes, past-light-cone construction.  The black continuous
    line gives ideal scaling, i.e. a constant time for the weak test
    and $N\log N$ for the strong test.}
\end{figure*}

Figure~\ref{fig:scalingtest} gives strong and weak scaling tests for the
new version of the code, obtained (analogously to M13)
by distributing a volume of 1.2 Gpc/h sampled by $900^3$ particles on
a number of cores ranging from $16$ to $14\times16=224$ (strong
scaling), and running a set of boxes, at the same mass resolution,
from $900^3$ to $2160^3$ particles (weak scaling).  In this test we
used 2LPT displacements for haloes building and catalogues, and we gave
enough memory so that fragmentation does not require the box to be
divided into slices (see point (2) above).  We show the total CPU time
(in hours) needed by several parts of the code.  An ideal scaling,
constant for the strong test and growing like $N\log N$ for the weak
test, is given by the black continuous line.  Because, thanks to the
FFTW package, \citep{frigo2012}, FFTs scale very closely to the
ideal case, and because the rest of the code performs distributed
computations and i/o is kept to a minimum, the scaling is very close
to ideal for the strong test and better than ideal for the weak test.

As a matter of fact, the FFT solver gives the strongest limitation to
the largest run that is feasible with this code. This is due to the
fact that memory is distributed in planes, so a task must have memory
for at least one plane. A mixed MPI-OpenMP configuration helps in
making this limitation less stringent: a plane is loaded onto an MPI
task that accesses more memory, the computation is distributed over
many threads using OpenMP. We have implemented this configuration and
verified that its scaling is worse than the pure MPI case.  We are
currently working to overtake this limit by using an FFT solver that
distributes memory with a more flexible geometry (in pencils or
cubes).

To understand whether the results presented in the paper depend on
resolution, for the same configuration used throughout the paper we
have run {\pin} using $512^3$ and $2048^3$
particles\footnote{This test was performed during the
    calibration phase. The parameters used in the runs adopted for
    this resolution test are not the final parameters presented in the
    paper, although they do not differ much. We have not re-run this
    test because the relative difference is what is relevant here.}.
The generation of random seeds allows to have the
same large-scale structure when the number of particles per side is
increased or decreased by factors of two.  In this test we use 2LPT
for both halo construction and displacement. From these
  runs, catalogues have been extracted with mass cuts of 100 and 500
  particles, as explained in Section~\ref{sect: catalogues}. The
  reference catalogue is that cut at 500 particles, in order to have a
  good sampling even in the lower resolution run.  In Fig. \ref{fig:
  resolution test} the power spectrum in real space is shown in the
upper panel, while the lower panel gives the residuals with respect to
the N-body power spectrum. When comparing the standard run with the
lower resolution one, {\pin} catalogues show some dependence on
resolution, but the high resolution run shows that convergence is
reached at our standard resolution.

\begin{figure}
  \centering
  \includegraphics[width=\columnwidth]{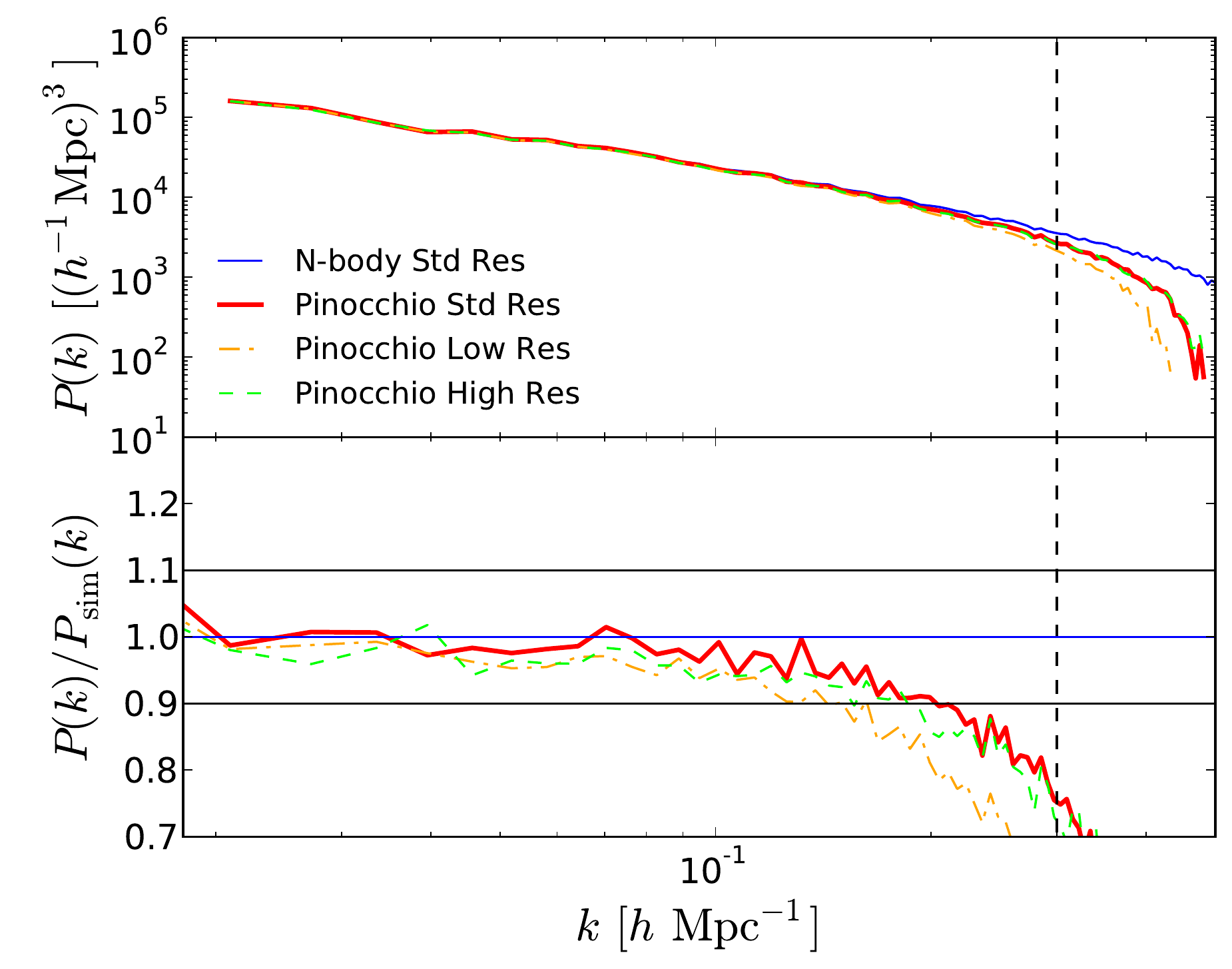}
  \caption{\label{fig: resolution test} Power spectrum in real space
    (\emph{top panel}) for different runs, both for N--body simulation
    and for {\pin}, as indicated in the legend: ``Std'' refers to
    the $1024^3$ particle runs, ``Low Res'' to the $512^3$ particle
    runs and ``High Res'' to the run with $2048^3$ particles and same
    box size. All the runs are made with 2LPT both for the halo
    construction and displacement. In the \emph{bottom panel} we show
    the ratio of the power spectra with that of the standard
    resolution N--body run. The horizontal blue line gives unity
    value, while the horizontal dotted ones give the $\pm10$ per cent
    accuracy region. The vertical dashed line gives $k_{MAX}$.}
\end{figure}

\section{Calibration}
\label{app:calib}

\begin{table}
\centering
\begin{tabular}{lcccc}
\hline
Name & Box size & N. & Particle Mass & N.\\
& (Mpc/h)  & particles & (M$_\odot/h$) & realizations\\
\hline
{\it VeLar} & $4096.0$ & $1024^3$ & $4.44\cdot 10^{12}$ & 10 \\
{\it Large} & $2048.0$ & $1024^3$ & $5.55\cdot 10^{11}$ & 10 \\
{\it Mediu} & $1024.0$ & $1024^3$ & $6.93\cdot 10^{10}$ & 10 \\
{\it Small} & $512.0$ & $1024^3$ & $8.67\cdot 10^{9}$   & 10 \\
{\it VeSma} & $256.0$ & $1024^3$ & $1.08\cdot 10^{9}$   & 10 \\
\end{tabular}
\caption{Main properties of the five sets of runs used for the calibration.}
\label{table:calibruns}
\end{table}

To calibrate {\pin} against an analytical, universal mass function, we
produced five sets of runs using $1024^3$ particles and box sizes from
$4096$ to $256$ Mpc$/h$ (Table~\ref{table:calibruns}). Particle masses
thus range from $1.08\cdot 10^{9}$ to $4.44\cdot 10^{12}$ M$_\odot/h$,
a factor of $\sim4000$ in variation. This allows to reliably sample
the mass function on almost five orders of magnitude in mass. For each
set we produced runs with 10 different random seeds, to beat down
sample variance. The {\it Mediu} boxes are analgous to the $1024^3$
setup of the paper, the first one being exactly the same run.

\begin{figure}
  \centering{\includegraphics[width=\columnwidth]{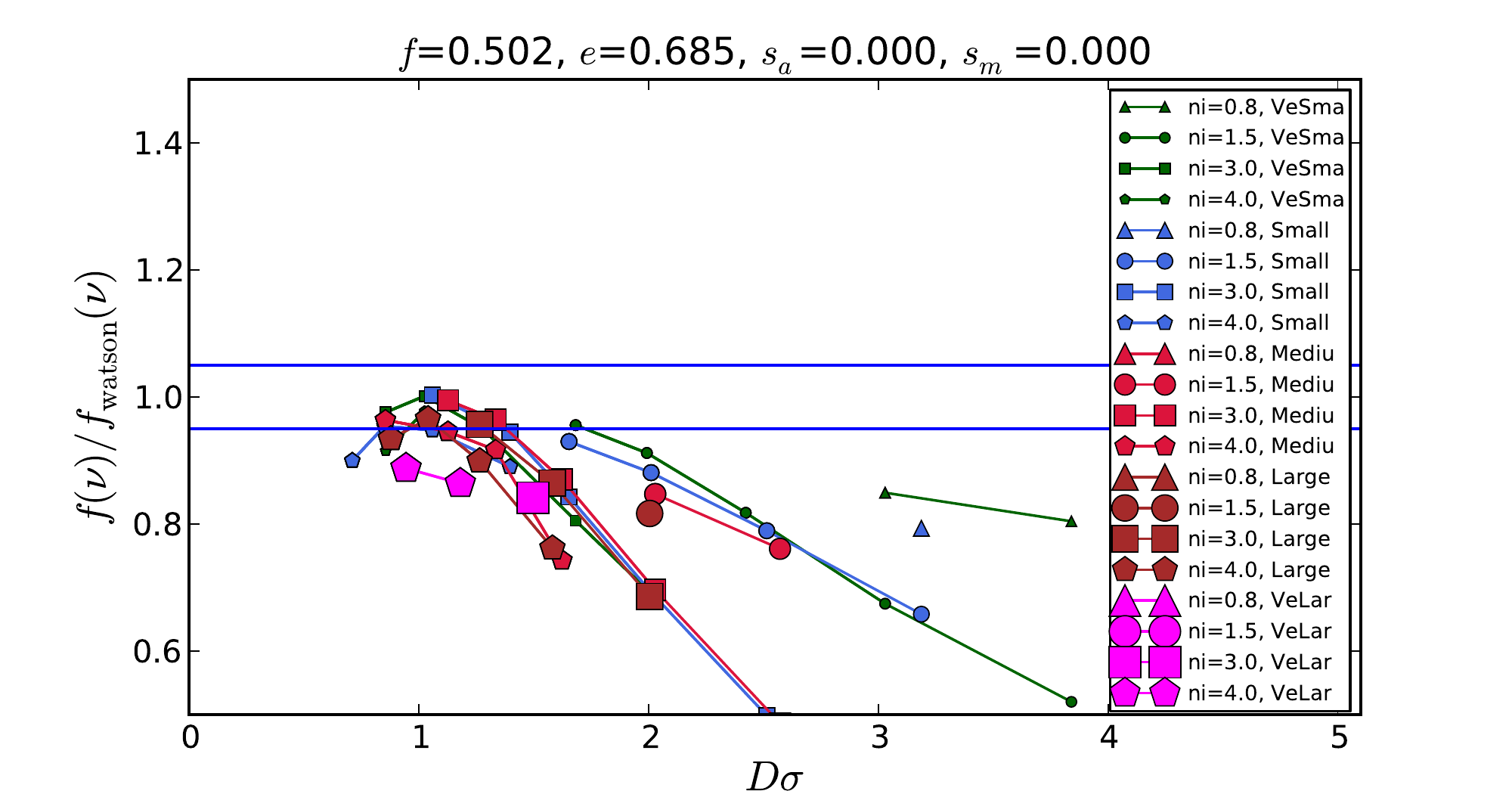}}
  \centering{\includegraphics[width=\columnwidth]{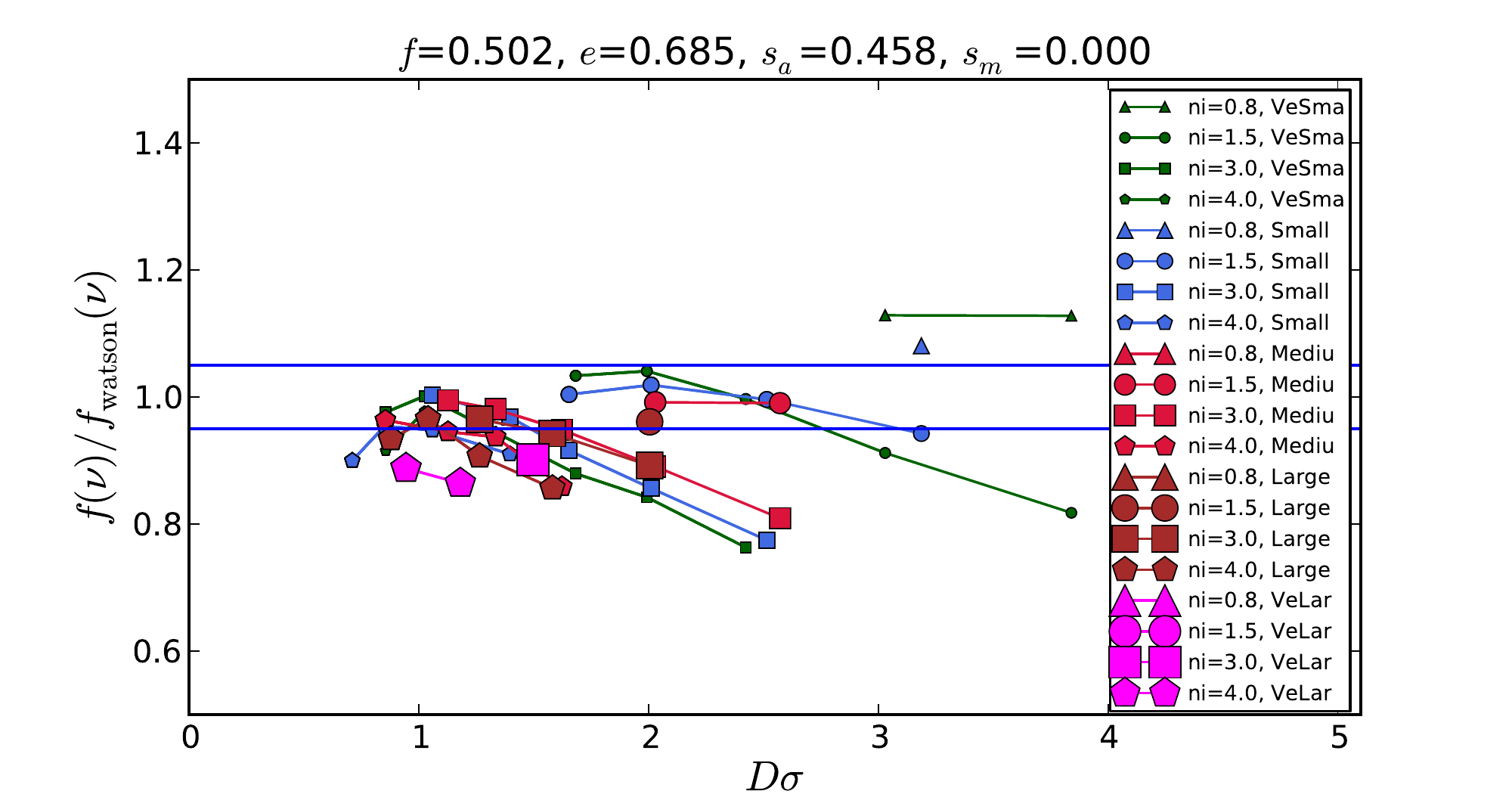}}
  \centering{\includegraphics[width=\columnwidth]{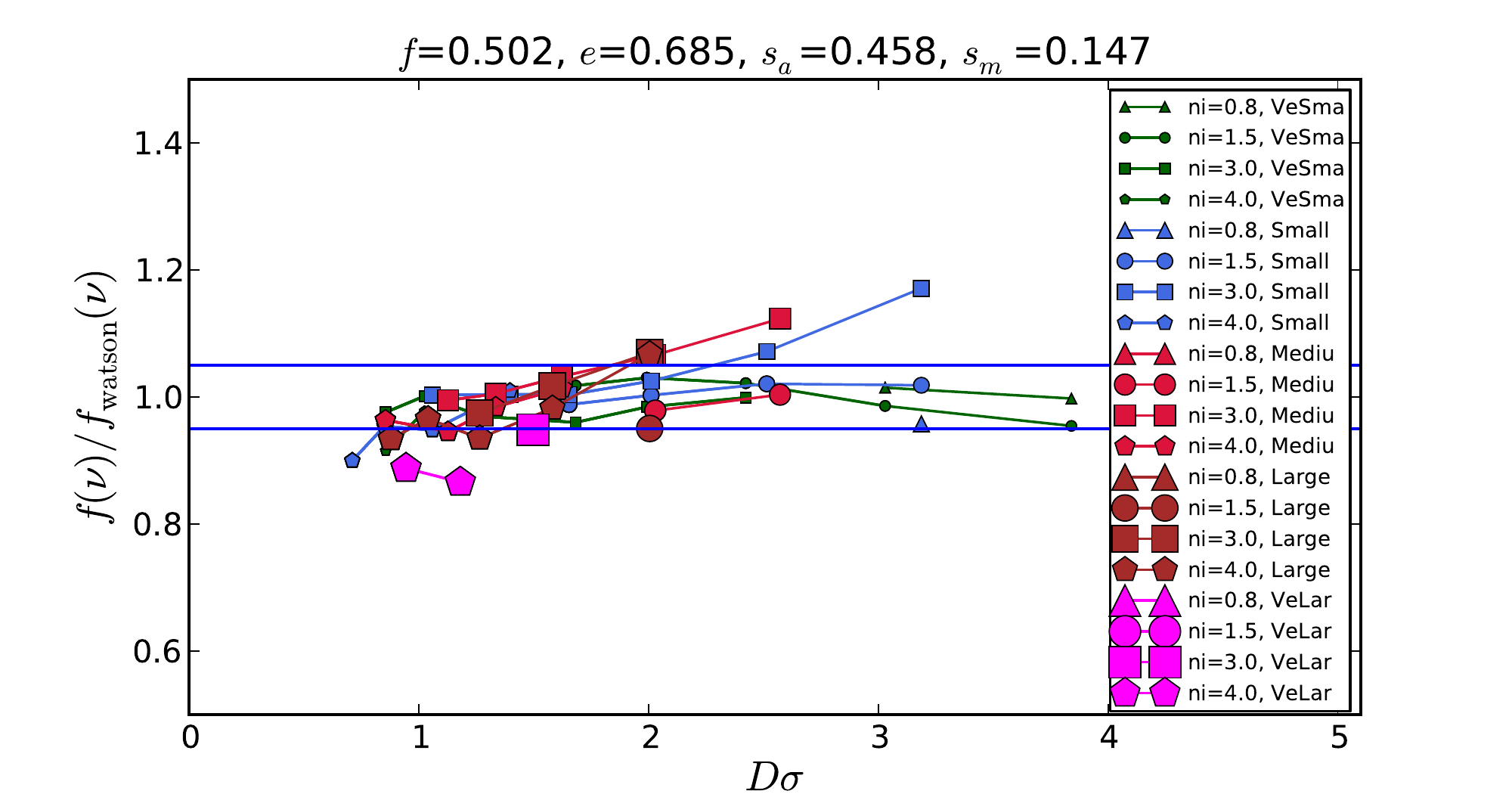}}
  \caption{\label{fig:universality} }
\end{figure}

To better visualize the violation of universality versus the level of
non-linearity $D\sigma$, we consider the rescaled mass function
$f(\nu)$ produced by {\pin} at some redshift $z$, corresponding to a
given $D\sigma$: $f(\nu,D\sigma)$. We take for each set of runs the
average over the ten realizations. We then take the ratio of this
quantity with an analytic, universal mass function; we consider here
the fit proposed by \cite{watson2013}. Then we consider four small
intervals of $\nu$ around four specific values, with semi-amplitude of
0.2, and compute the average value of $f(\nu,D\sigma)/f_{\rm
  watson}(\nu)$ in the bin. This is a set of functions of $D\sigma$,
one for each set of runs and each $\nu$ value. The upper panel of
Figure~\ref{fig:universality} shows these functions for 3LPT
displacements and a choice of parameters $f=0.501$, $e=0.745$, while
$s_a=s_e=0$.

The functions show a clear maximum at $D\sigma\sim1.2$. We take this
as the value of the $D\sigma_0$ parameter. Results change very slowly
with $D\sigma_0$, and small variations are degenerate with $s_a$ and
$s_m$, so there is no need to fine-tune this parameter. At lower
values, {\pin} mass functions show a modest systematic underestimate with
respect to Watson. We noticed a similar behaviour with simulations, so
we interpret this as a sign that mass resolution is not sufficient to
properly reproduce haloes and do not attempt to correct for it. On the
other hand, the decrease at $D\sigma>D\sigma_0$, that is not seen in
the N-body case (where the curves grow due to the violation of
universality), is interpreted as the effect of increasing inaccuracy
of displacements. The lower panels show the effect of using optimal
values of $s_a=0.334$ and $s_m=0.052$ to improve the fit.

\begin{figure}
  \centering
  \includegraphics[width=\columnwidth]{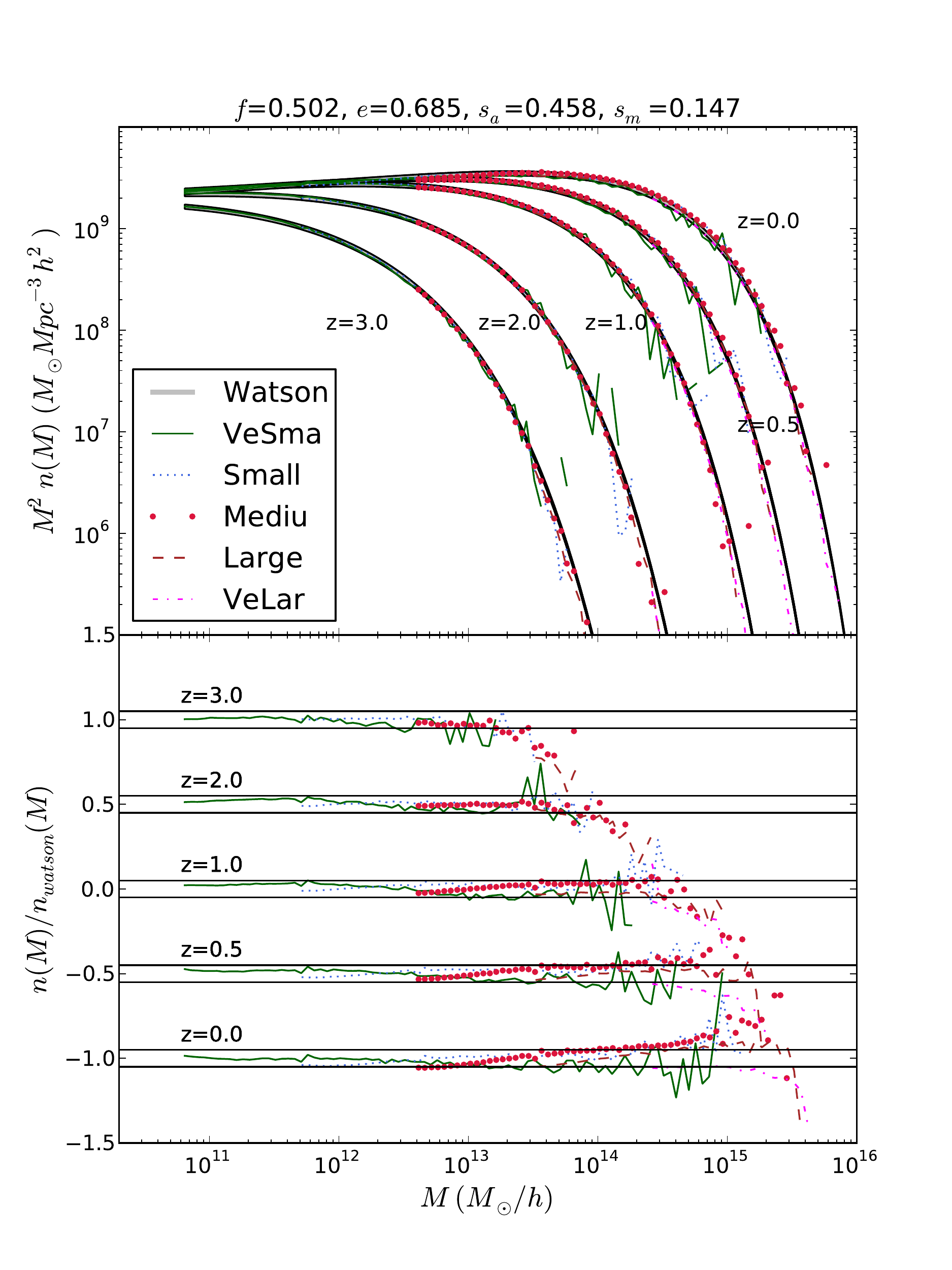}
  \caption{\label{fig:MFbest}Mass function produced for the five sets
    of runs, compared with the Watson et al., 2013 fit. The upper
    panel shows the quantity $M^2n(M)$ as a function of $M$ The lower
    panel shows the residuals with respect to the analytic fit,
    conveniently displaced. The horizontal black lines give the $\pm 5$
    per cent region.}
\end{figure}

Best-fit parameters were found by running the code on a grid of
parameter values, studying the effect of their variations on the mass
function at some relevant $\nu$ and $D\sigma$ values, finding the
degeneracies between parameters, and then guessing a value that
minimizes the differences. Figure~\ref{fig:MFbest} shows the final
calibrated mass function. As in Figure~\ref{fig:calibratedMF}, the
upper panel shows $M^2n(M)$ for the five sets of runs, together with
the Watson analytic mass function, while the lower panels show the
residuals. 
We will show below that the steepening of some mass functions at low
$z$ is within the uncertainty of the numerical mass function.
In principle one could recalibrate the code for any
analytic mass function. At fixed $D\sigma$, some very small variations
with resolution, at the few percent level, are present in the figures.
They grow to $\sim10$ percent at the lowest resolution {\it VeLar}. To
fix it one could introduce an explicit dependence on resolution
(through $\sigma$ in place of $D\sigma$) but the mass resolution of
{\it VeLar} is so poor that we do not foresee any application of it.

\begin{figure}
  \centering{\includegraphics[width=\columnwidth]{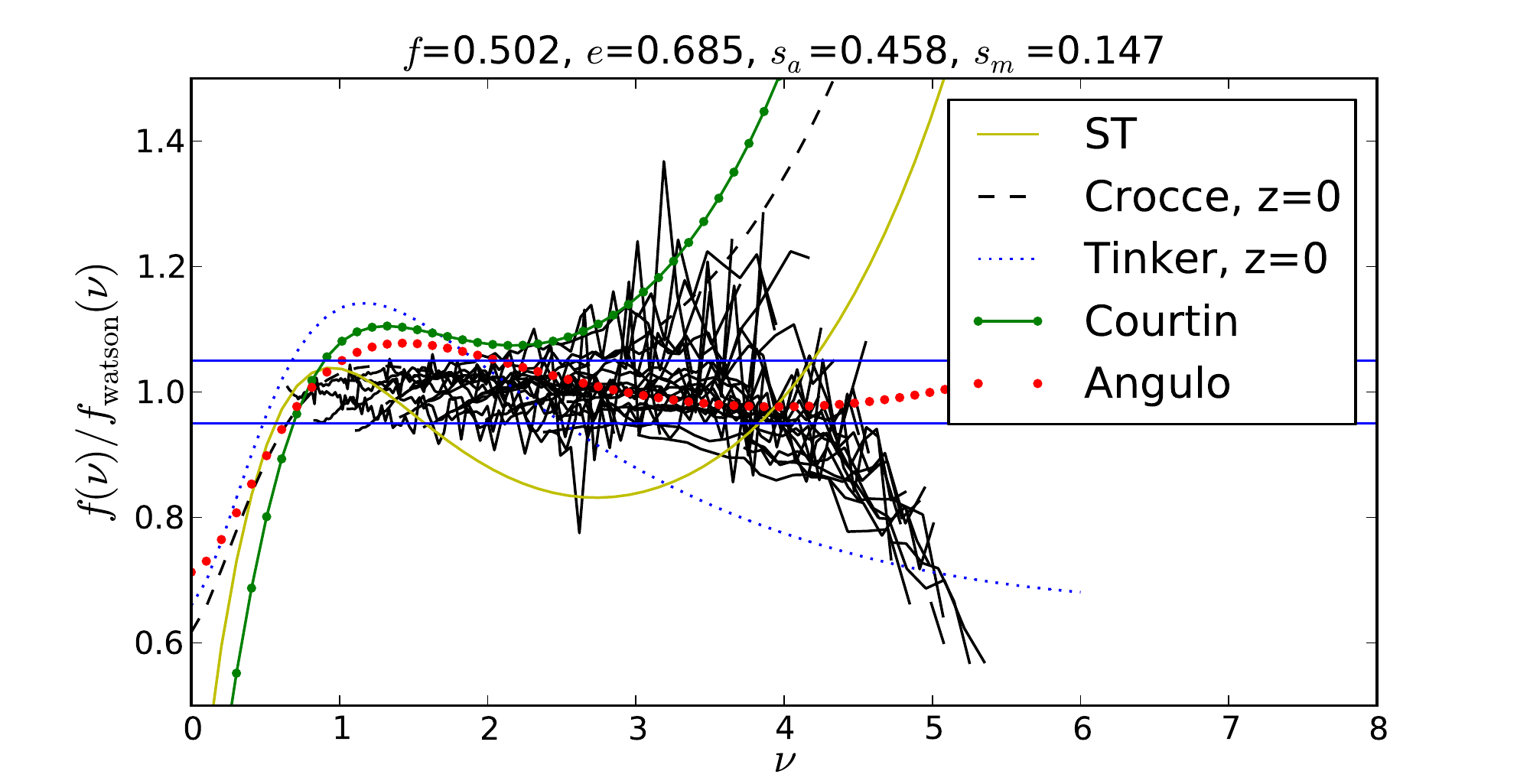}}
  \caption{\label{fig:fnu} Rescaled mass function $f(\nu)$, divided by
    the Watson et al., 2013 fit, for all the runs (black lines)
    performed with 2LPT and for the best-fit parameters. Colored
    curves give other numerical fits, as illustrated in the legend
    (see references in the text).}
\end{figure}

The high tail of the mass function ($\nu\ge3$) tends to overestimate
the Watson fit at late times, with some dependency on resolution. In
figure~\ref{fig:fnu} we show the bunch of $f(\nu)$ curves versus
several analytic formulas, including \cite{sheth2002},
\cite{crocce2010}, \cite{tinker2008}, \cite{courtin2011} and
\cite{Angulo2012}. At $\nu\sim3-4$ mass functions predicted by
{\pin} show a spread, filling the region from the Watson and Angulo
fits to the Crocce and Courtin one. This trend grows with time, and is
stronger at lower resolution (higher $D\sigma$). The likely origin of
this trend is the following. The separation of merging and accretion
naturally depends on mass resolution, as a particle that accretes on a
halo will possibly contain haloes at a lower resolution. We are fixing
the growing inaccuracy of displacements by increasing merging and
accretion in a separate way, through constant $s_a$ and $s_m$
parameters. A more sophisticated approach should take into account
the resolution-dependent separation of accretion and merging.
Because the results are within the uncertainty of the numerical mass
function, we do not attempt to implement this sophistication now.

Analogous calibrations have been performed for ZA and 2LPT. Best-fit
parameters are given in Table~\ref{table:bestanalytic}. All these
parameters give a universal mass function. In this context,
non-universality can be easily obtained (as in the main text to follow
the trend of the N-body simulation) at a given resolution but, because
the main dependency is with $D\sigma$, the redshift evolution of this
non-universality will depend on resolution.

\begin{table}
\centering
\begin{tabular}{lrrr}
\hline
Parameter & ZA value & 2LPT value & 3LPT value\\
\hline
$f$         &  0.505 &  0.501 &  0.502\\
$e$         &  0.820 &  0.745 &  0.685\\
$s_a$       &  0.300 &  0.334 &  0.458\\
$s_m$       &  0.000 &  0.052 &  0.148\\
$D\sigma_0$ &  1.7   &  1.5 &  1.2 \\
\hline \hline 
\end{tabular}
\caption{\label{table:bestanalytic} Adopted values of the parameters of
  eq. \ref{eq:calib3} for the calibration of the mass function against
  the Watson et al., 2013 analytical fit.}
\end{table}




\bibliographystyle{mnras}
\bibliography{bibliography} 


\bsp 
\label{lastpage}
\end{document}